\documentclass[journal]{IEEEtran}
\usepackage{amsthm}
\usepackage{amsfonts}
\usepackage{gensymb}
\usepackage{amssymb}
\usepackage[export]{adjustbox}
\usepackage{pifont}
\usepackage{siunitx}
\usepackage{graphicx}
\usepackage{array}
\usepackage{lscape}

\usepackage{caption}
\usepackage{subcaption}
\usepackage{multicol}
\usepackage[cmex10]{amsmath}
\usepackage{algorithm}
\usepackage{algorithmicx}
\def\algbackskip{\hskip-\ALG@thistlm}
\usepackage{flushend}

\usepackage{relsize}
\usepackage{hyperref}
\usepackage{xcolor}
\usepackage{tikz}
\usepackage{algpseudocode}
\usepackage[justification=centering]{caption}

\newtheorem*{remark}{Remark}
\newcommand{\Rom}[1]{\uppercase\expandafter{\romannumeral#1\relax}}

  \makeatletter
\def\endthebibliography{%
  \def\@noitemerr{\@latex@warning{Empty `thebibliography' environment}}%
  \endlist
}
\makeatother
\begin{document}
\title{Joint Angle and Velocity-Estimation for Target Localization in Bistatic mmWave MIMO Radar in the Presence of Clutter}
\author{Priyanka~Maity,~\IEEEmembership{Graduate Student Member,~IEEE,}  Suraj~Srivastava,~\IEEEmembership{ Member,~IEEE,} Aditya~K.~Jagannatham,~\IEEEmembership{Senior Member,~IEEE.}  and Lajos~Hanzo,~\IEEEmembership{Life Fellow,~IEEE} %
\thanks{L. Hanzo would like to acknowledge the financial support of the Engineering and Physical Sciences Research Council (EPSRC) projects under grant EP/Y037243/1, EP/W016605/1, EP/X01228X/1, EP/Y026721/1, EP/W032635/1, EP/Y037243/1 and EP/X04047X/1 as well as of the European Research Council's Advanced Fellow Grant QuantCom (Grant No. 789028). The work of A. K. Jagannatham was supported in part by the Qualcomm Innovation Fellowship; in part by the Qualcomm 6G UR Gift; and in part by the Arun Kumar Chair Professorship. The work of S. Srivastava was supported in part by IIT Jodhpur's Research Grant No. I/RIG/SUS/20240043; in part by Anusandhan National Research Foundation's PM-ECRG/2024/478/ENS; and in part by Telecom Technology Development Fund (TTDF) under Grant TTDF/6G/368. S. Srivastava and A. K. Jagannatham jointly acknowledge the funding support provided to ICON-project by DST and UKRI-EPSRC under India-UK Joint opportunity in Telecommunications Research.}
  \thanks{P. Maity and A. K. Jagannatham are with the
Department of Electrical Engineering, Indian Institute of Technology,
Kanpur, Kanpur, 208016, India (e-mail: pmaity@iitk.ac.in,  
adityaj@iitk.ac.in.)}%

\thanks{S. Srivastava is with the Department of Electrical Engineering, Indian Institute of Technology, Jodhpur, Jodhpur, 342030, India (e-mail:
surajsri@iitj.ac.in.) }%

\thanks{L. Hanzo is with the School of Electronics and Computer
Science, University of Southampton, Southampton SO17 1BJ, U.K.
(e-mail: lh@ecs.soton.ac.uk)} 
} 
\maketitle
\begin{abstract}
Sparse Bayesian learning (SBL)-aided target localization is conceived for a bistatic mmWave MIMO radar system in the presence of unknown clutter, followed by the development of an angle-Doppler (AD)-domain representation of the target-plus-clutter echo model for accurate target parameter estimation. The proposed algorithm exploits the three-dimensional (3D) sparsity arising in the AD domain of the scattering scene and employs the powerful SBL framework for the estimation of target parameters, such as the angle-of-departure (AoD), angle-of-arrival (AoA) and velocity. To handle a practical scenario where the actual target parameters typically deviate from their finite-resolution grid, a super-resolution-based improved off-grid SBL framework is developed for recursively updating the parameter grid, thereby progressively refining the estimates. We also determine the Cramér-Rao bound (CRB) and Bayesian CRB for target parameter estimation in order to benchmark the estimation performance. Our simulation results corroborate the superior performance of the proposed approach in comparison to the existing algorithms, and also their ability to approach the bounds derived.
\end{abstract}
\IEEEpeerreviewmaketitle
\section{Introduction}
\textcolor{black}{Next-generation (NG) wireless systems are expected to support high-accuracy localization services in indoors for robot navigation \cite{8365918}, human sensing \cite{9941043}, Wi-Fi sensing for smart homes \cite{8999605}, radar sensing for autonomous vehicles \cite{bilik2019rise}, location-assisted communication \cite{6924849, 9782674}, user positioning \cite{9774917, 9364875} and location-aware applications \cite{7462484, 8847221}, among others. Estimation of distance and angle is also crucial for distance-aware resource allocation and precoding in NG wireless systems \cite{7979497}. The primary objective of localization is to identify the position and velocity of a target via estimating the direction of arrival, Doppler shift, and time delay.} As a further advance, multiple-input multiple-output (MIMO) radar promises significant diversity and multiplexing gains \cite{li2007parameter, haimovich2007mimo}. MIMO radar systems can be classified into colocated and distributed MIMO \cite{liu2022survey}, wherein the transmitter and receiver are at the same position in the former, while they are distributed in different locations in the latter.

Millimetre wave (mmWave) carriers, which belong to the extremely high frequency (EHF) range spanning from 30 GHz to 300 GHz, have extraordinarily short wavelength. This enables the packing of a large number of antennas in devices having compact form factor \cite{9050553}, which in turn enables ultra-high angular resolutions that make them ideally suited for radar operation \cite{6127923}. Millimetre wave (mmWave) technology is expected to become increasingly dominant in the near future due to the availability of sizeable unlicensed frequency blocks \cite{6515173}, which makes it eminently suitable for demanding applications like ultra-high-definition (UHD) 3D video, virtual
and augmented realities, internet-of-things (IoT), satellite
communications, unmanned aerial vehicles (UAV), V2X, among others \cite{7968418}. The MIMO technology \cite{8288677} relying on a large number of transmit and receive antennas, coupled with the huge bandwidth of the mmWave regime, has the potential of significantly improving radar performance.

A mmWave distributed MIMO radar, composed of multiple bistatic transmitter/ receiver pairs, offers several advantages over its colocated counterpart, such as spatial diversity, high resolution, reduced transmitter-receiver antenna interference, while dispensing with switches or duplexers \cite{8828030}. Exploiting the angular and temporal sparsity of the mmWave channel in such a system may lead to a significant improvement in target detection and localization performance \cite{6803957,9449071}. Therefore, we conceive techniques for joint parameter estimation in target localization of bistatic mmWave MIMO radar. An
overview of the existing contributions in this area is presented next.
\subsection{Importance of location-aware
Interaction in the mmWave Band}
A specific drawback of operating at high frequencies, such as in the mmWave band, is that propagation losses increase with the carrier frequency. Potential solutions to overcome such losses in 5G networks is to utilize accurate location information \cite{6924849}. Firstly, knowing the positions and thus the distances between nodes can help estimate received power and interference levels. In dense networks, the shortest-distance multihop path between a source and destination is often the most efficient. Additionally, distance and angle estimation helps in location-aware resource allocation \cite{7979497}. Traditional resource allocation relies heavily on real-time channel state information (CSI), which must be frequently updated through pilot signals and feedback leading to overhead and delay
However, location-aware systems can predict future channel conditions by using the known or estimated position and mobility pattern of a device \cite{10792908, 10430216}. Since the path loss, interference, and even shadowing can be correlated with position, the system can estimate the CSI ahead of time—especially in slowly varying environments or when user trajectories are predictable (e.g., vehicles on roads, users on foot). The angle and distance estimation also helps attain improved energy and spectral efficiency by steering beams accurately in directional mmWave communication and turning off the antennas or reduce transmit power when coverage is known to be sufficient \cite{9382000}. Location-aware precoding is also beneficial in near-field communications which must account for both distance and angle information \cite{10623508,9860745}.
\subsection{Literature Review} 

He \textit{et al.} \cite{5765430} proposed a two-dimensional multiple signal classification
 (2D-MUSIC) algorithm for joint angle-of-arrival (AoA) and angle-of-departure (AoD) estimation in MIMO radar that imposes high computational cost. To reduce this, Zhang \textit{et al.} \cite{zhang2010direction} developed a reduced-dimensional MUSIC (RD-MUSIC) algorithm that searches for peaks over a single dimension. Another novel subspace-based method disseminated by Zheng \textit{et al.} \cite{7731139} is the estimation of signal parameters via rotational invariance
 techniques (ESPRIT) algorithm that is significantly faster than the RD-MUSIC since a closed-form angle estimation solution is obtained by exploiting the invariance property between the transmit and receive arrays. Nevertheless, their algorithm is unable to achieve the same level of accuracy as the family of MUSIC-related techniques. For precisely determining the signal/noise subspaces, both the ESPRIT and MUSIC-based approaches necessitate an excessively large number of snapshots, whose accuracy significantly degrades in scenarios of low signal-to-noise ratios (SNR) and/ or insufficient snapshots. With objective of joint AoD and AoA estimation, Tang \textit{et al.} \cite{tang2013maximum} developed an innovative maximum likelihood scheme, wherein an alternating projection algorithm was devised for resolving the associated high-dimensional nonlinear optimization. While it yields improved performance, its convergence is not guaranteed. The treatises \cite{qin2018efficient}, \cite{7468562} predominantly focus on the localization of targets in distributed radar based on time-of-arrival (ToA) and time difference of arrival (TDOA) measurements, respectively. Qin \textit{et al.} \cite{qin2018efficient} exploit the bistatic range (BR) and TDOA measurements for developing an interesting two-step least squares algorithm for target localization. By contrast Nguyen \textit{et al.} \cite{7468562} conceive a ToA based geometric model for minimizing the estimation confidence area by maximizing the determinant of the pertinent Fisher information matrix (FIM).
 However, the above contributions do not include velocity or Doppler shift estimation in the localization problem. As a further advance, the authors of \cite{9197719}, \cite{9099825} designed potent schemes for location and Doppler estimation based on BR, TDOA and Doppler shift measurements. Nevertheless, it is imperative to mention that TOA-based schemes require accurate time synchronization between the transmitters and receivers, which is challenging to achieve in practical systems. Furthermore, the scattering environment at mmWave frequencies is inherently sparse, a feature that can be exploited for bolstering performance, and one which none of the aforementioned studies fully exploit. The literature specifically related to compressive sensing (CS)-based localization is reviewed next.
 
 CS-based methods exhibit several key advantages over both the conventional signal processing methods and over the existing subspace based schemes. For example, they require a remarkably reduced number of snapshots for reliable signal recovery even for an unknown number of sources. Furthermore, they exhibit substantially improved robustness to noise \cite{malioutov2005sparse}.
Chen \textit{et al.} \cite{8537983} developed a sparse Bayesian learning (SBL)-based technique for direction of arrival (DOA) estimation in colocated MIMO radar. The joint estimation of the AoA and AoD parameters of point targets in MIMO-aided radar systems using SBL was successfully carried out in \cite{8379194}. As a further developement, Maity \textit{et al.} \cite{10683028} proposed an SBL scheme to jointly estimate the AoA, range and velocity of the targets in mmWave MIMO radar systems. Traditional SBL approaches require the targets to lie on a predefined grid of the AoD/AoA parameters. This in turn mandates the fixed parameter grid to be finely spaced in order for the sparse assumption to be reliable, which could lead to a high computational cost and violate the restricted isometric property (RIP) to be satisfied for sparse recovery \cite{RIPProperty}. The pioneering work by Yang \textit{et al.} \cite{offGrid1} addressed the grid mismatch problem by proposing an off-grid model obtained by linearly approximating the dictionary and updating it in every iteration. However, it does not fully eliminate the unavoidable off-grid performance erosion. To address this impediment, Cao \textit{et al.} \cite{8880558} jointly estimated the AoDs and AoAs in bistatic MIMO radar by introducing a modified linear approximation model, where the grid points are directly updated in every iteration for ensuring that the updated estimates approach the true AoAs and AoDs upon convergence. However, the exhaustive list of papers discussed above ignore the mobile targets and ground clutter. The undesired echo imposed by the ground clutter such as land, buildings, roads, etc \cite{shnidman1999generalized, billingsley1999statistical} further degrade the sensing performance, if not mitigated. This limits the localization performance in practical integrated sensing and communication (ISAC) systems.
\par
Several clutter suppression techniques \cite{luo2024integrated, sun2020structured, zhang2019reduced} identify and eliminate clutter by relying on time-domain and space-time-domain approaches. However, these do not exploit the angular domain sparsity. Mishra \textit{et al.} \cite{8529199} leveraged the target sparsity and estimated the target parameters using the SBL framework in the presence of ground clutter for colocated radar systems. 
\begin{table*}
\centering
\caption{Boldly contrasting our new contributions against the state-of-the-art}
\begin{adjustbox}{width=1\textwidth}
        \begin{tabular}{ |l | c | c |c |c |c |c |c |c |c |c |c |c |c |c |c |c | }
        \hline
            \textbf{Features} & \cite{8365918} & \cite{5765430,zhang2010direction, tang2013maximum} &  
  \cite{7468562} & \cite{9099825}  & \cite{8379194} &\cite{10683028} & \cite{offGrid1} & \cite{8880558} & \cite{8529199}  & \textbf{Proposed}\\
            \hline
            mmWave band & & & & & &\checkmark & & & &  \checkmark\\
            \hline
            Distributed radar &\checkmark &\checkmark & \checkmark  & \checkmark  & \checkmark & & & \checkmark&   &    \checkmark\\
            \hline
            Off-grid SBL & & & & & \checkmark & &   \checkmark&   \checkmark &  &   \checkmark \\
            \hline
            Multiple targets &  & \checkmark & & & \checkmark & &   \checkmark &   \checkmark& & \checkmark \\
            \hline
            Multiple parameter estimation & & $\boldsymbol{\theta, \phi}$ & ${\phi}$ & $\mathbf{p},v$  & $\boldsymbol{\theta}$ & RCS, $\theta, v$ & $\boldsymbol{\theta}$ & $\boldsymbol{\theta, \phi}$ & RCS, $\theta, v$  & RCS, $\theta, \phi, v$ \\
            \hline
            Target localization & \checkmark &  & \checkmark & \checkmark &  & \checkmark & & &  &   \checkmark\\
            \hline
            CRB for joint parameter estimation&  & & & & & & & & \checkmark  &  \checkmark\\
            \hline
            3D sparsity & & & & & &\checkmark &\checkmark &\checkmark &\checkmark  &  \checkmark\\
        \hline
                   Unknown clutter & \checkmark  & & & & &\checkmark & & & &  \checkmark\\
            \hline
            Geometric calculation of location from AoA and  AoD&  &  & & & & & & & &   \checkmark\\
            \hline            
        \end{tabular}
        \end{adjustbox}
        \label{Tab:Comparision}
\end{table*}
However, they assumed the clutter covariance matrix to be known, which is challenging to estimate in real-world systems. To fill these research gaps, this treatise conceives a Bayesian learning-based localization scheme that is capable of jointly estimating the AoA, AoD, velocity and radar cross-section (RCS) parameters of multiple targets in the presence of clutter having unknown statistics, while exploiting the 3D sparsity arising in the angular and Doppler (AD)-domain. The main contributions of this study are itemized next and they are also boldly contrasted to the literature in Table-\ref{Tab:Comparision}. The symbols ${\phi,\theta}, v$ used in the table denote the target AoA, AoD and velocity, respectively. \par
\subsection{Contributions of our work}
\textcolor{black}{ \begin{itemize}
\item A novel bistatic mmWave MIMO radar system is developed for the localization of multiple mobile targets and joint estimation of target parameters such as AoA, AoD, Doppler shift and RCS coefficients in the presence of unknown noise-plus-clutter components.
\item The localization problem is viewed from a fresh perspective of
the emerging field of compressive sensing (CS) that exploits the $3$D sparsity in the angle-Doppler (AD)-domain. To overcome the drawback of uniform sampling of the grids toward dictionary matrix formulation, this work presents a framework for non-uniform sampling of the 3D grid, where no two grid points share the same AoA, AoD and Doppler
shift values.
\item As a result of non-uniform sampling, modeling error arises due to the gap between the true target parameter and its closest grid point. In order to address the off-grid problem, the adjustable grid points are updated recursively, until convergence is reached.
Subsequently, a novel block majorization-minimization (MM) procedure is developed for Bayesian inference.
\item Furthermore, the proposed algorithm helps filter out the unknown clutter components from the estimated target components. Subsequently, a localization algorithm is devised for determining the position coordinates and velocity of the targets from the estimated AoAs, AoDs and Doppler shifts.
\item Next, the Cram\'er Rao bounds (CRB) are determined for joint AoA, AoD and velocity vector estimation, which benchmark the estimation performance of the proposed design. Furthermore, the Bayesian CRB (BCRB) is derived for RCS coefficient estimation in order to characterize its estimation performance.
\item Lastly, our simulation findings comprehensively validate the performance of the proposed schemes and clearly evidence  their enhanced performance over the existing methods, even at low signal to clutter-plus-noise ratio (SCNR).
\end{itemize}}
The rest of the paper is outlined as follows, The system model of bistatic mmWave MIMO radar is developed in Section \ref{system_model} in presence of clutter. Section \ref{BL} discusses the off-grid SBL-based algorithm conceived for joint AoA-AoD-Doppler shift estimation, followed by our localization algorithm in Section \ref{localization} and complexity analysis in Section \ref{complexity}. Section \ref{crb} derives the CRB whereas our simulation results are discussed in Section \ref{results}. 
Section \ref{conclusion} concludes the paper.
\subsection{Notation} 
 The operator $\Re\big\{.\big\}$ denotes the real part of a complex-valued argument, while $\vert\mathbf{a}\vert$,   $\vert\vert\mathbf{a}\vert\vert_0$ and $\vert\vert\mathbf{a}\vert\vert_2$ denote the element-wise absolute value, the $\ell_0$ norm and the Euclidean norm of the vector $\mathbf{a}$, respectively. For a matrix $\mathbf{A}$, $\vert\mathbf{A}\vert$, $\text{Tr}(.)$ and $\vert\vert\mathbf{A}\vert\vert_F$ represent its determinant, trace and the Frobenius norm, respectively. The notations rect(.) and $\mathrm{sign}(.)$ stand for the rectangular pulse and signum function, respectively. Furthermore, $\mathbf{1}_N$ and $\mathbf{I}_N$ represent an $N \times 1$ vector with entries as 1 and $N \times N$ identity matrix, respectively. The operator diag(.) results in a diagonal matrix having the elements of its argument vector on the principal diagonal. Operator vec(.) performs vectorization of its argument matrix by stacking its columns. The operators $(.)^H, (.)^*, (.)^T$ denote the Hermitian, conjugate and transpose operation applied to the argument, respectively.
 The
operations $A\odot B$ and $A \otimes B$ denote the Hadamard and Kronecker matrix
products, respectively. The operator $\mathbb{E}$(.) represents the statistical expectation, while $\int(.) dw$ and $\frac{\partial}{\partial w}(.)$ are the integration and partial derivative of the argument with respect to $w$.
\section{System Model for bistatic mmWave MIMO radar} 
\subsection{Signal Model}
\begin{figure*}[h]
\centering\includegraphics[scale=0.55]{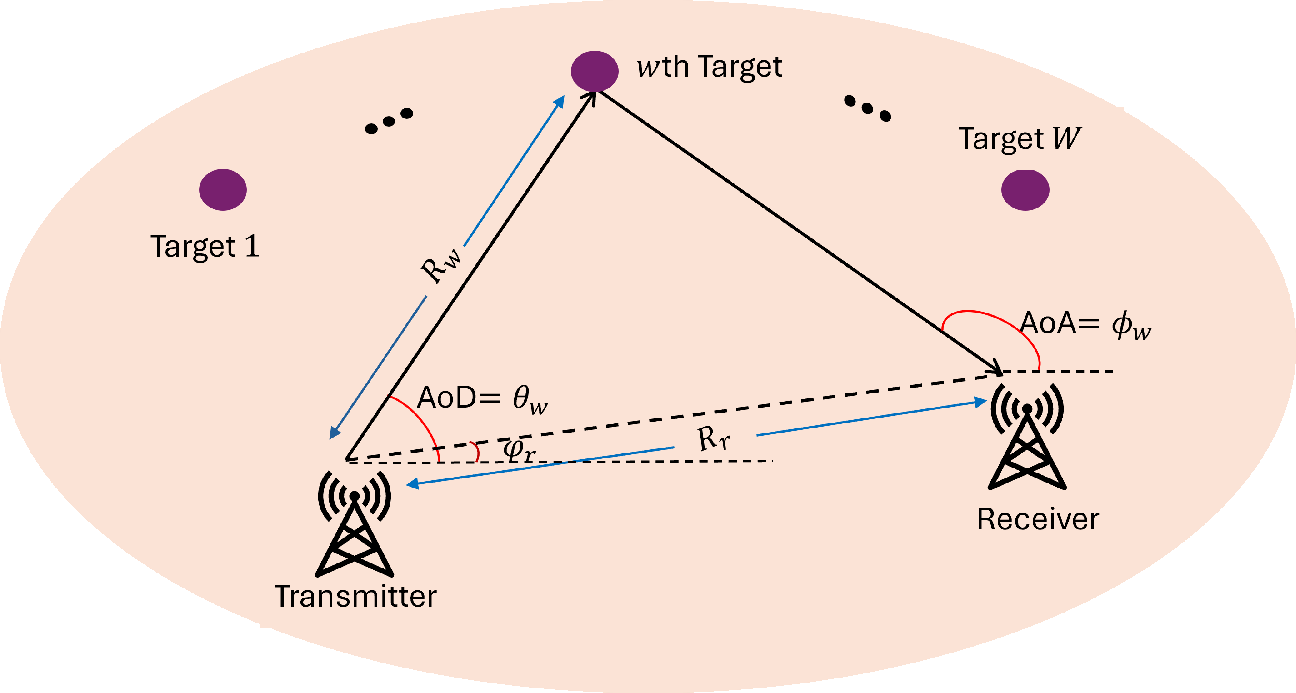}
\caption{\textcolor{black}{System model for target localization in bistatic mmWave MIMO radar}}
\label{system_model}
\end{figure*}

A bistatic mmWave MIMO radar setup comprising a transmitter and a receiver is considered, which are equipped with $N_t$ transmit and $N_r$ receive antennas, respectively,  as shown in Fig. \ref{system_model}.
The transmitter is situated at the origin and the receiver's coordinates are given $\left[R_r,\varphi_r\right] \in \mathbb{R}^2$, where $R_r$ is the distance between the receiver and transmitter, while $\varphi_r$ is the orientation. The $w$th target's unknown polar coordinates are $\left[R_w,\varphi_w\right] \in \mathbb{R}^2$. Note that this model can be easily extended to a $3$D geometry, where the $w$th target's unknown polar coordinates can be identified as $\left[R_w, \theta_w, \phi_w \right]$. Here, $\theta_w\ \text{and}\  \phi_w $ represent the azimuth and elevation angle, respectively. Furthermore, the proposed localization framework can be easily adapted to a 3D scenario where the sparsity in the azimuth angle-elevation angle-Doppler domain will be leveraged. For simplicity, the study is limited to the $2$D geometry. Let $W$ denote the number of unknown moving targets present in the network and let the velocity of the $w$th target be $v_w$. The signal impinging from the $t$th transmit antenna, $ 1 \leq t\leq N_t$, can be represented as
\begin{equation}
   \mathbf{{ {x}} }_t^T =  
\begin{bmatrix}
x_t(1), & x_t(2), &\ldots, & x_t(U)
\end{bmatrix} \in \mathbb{C}^{1\times U}, 
\end{equation}
where $U$ is the total number of sub-pulses. Thus, $\mathbf{X} =  \begin{bmatrix}\mathbf{{ {x}}}_1 &\mathbf{{ {x}}}_2 &\ldots &\mathbf{{ {x}}}_{N_t} \end{bmatrix}^T \in \mathbb{C}^{N_t\times U}$ denotes the overall transmit signal matrix. The Doppler-shifted signal corresponding to the $t$th antenna and $w$th target is defined as $\mathbf{{x}}_t^T(\omega_w) = \mathbf{{x}}_t^T\odot \boldsymbol{{\phi}}^T(\omega_w) $, where $\boldsymbol{{\phi}}^T\left(\omega_w\right)=  
\begin{bmatrix}
1, & e^{j\omega_w} &\ldots, & e^{j\omega_w (U-1)}
\end{bmatrix} \in \mathbb{C}^{1\times U}$ is a vector corresponding to the Doppler shift $\omega _w$ of the $w$th target, which is defined as 
\begin{align}
   \omega _w = 2\pi f_w T_p, \label{doppler_shift}
\end{align}
where $f_w = \frac{v_w}{\lambda}$ denotes the Doppler frequency and $T_p$ is the sub-pulse interval.
We define the matrix $\mathbf{X}_w= \Big[
\mathbf{x}_1(\omega_w),
\mathbf{x}_2(\omega_w),
\cdots,
\mathbf{x}_{N_t}(\omega_w) \Big]
^T\in \mathbb{C}^{N_t\times U}$, which represents the transmitted signal matrix associated with the motion of the $w$th target.  
\textcolor{black}{Upon considering $N_{cl}$ ground clutter components, the signal $\underline{\mathbf{Y}} \in \mathbb{C}^{N_r\times U}$ reflected from the $W$ targets and arriving at the receiver is modeled as
\begin{equation} 
\underline{\mathbf{Y}}= 
\sum_{w = 1}^{W} e^{\frac{-j2\pi R_{w}^B}{\lambda}} \beta_w\mathbf{ {d}}_{w}\mathbf{ {c}}^T_w\mathbf{X}_w + \sum_{z = 1}^{N_{cl}} e^{\frac{-j2\pi R_{z}^B}{\lambda}}\beta_{z}\mathbf{ {d}}_{z}\mathbf{ {c}}^T_{z}\mathbf{X} + \mathbf{Q}
\label{recv_node},
\end{equation}
where 
$\beta_w\in \mathbb{C}$ and $\beta_{z}\in \mathbb{C}$ are the RCS coefficients of the $w$th target and $z$th clutter, respectively. The factors $e^{\frac{-j2\pi R_{w}^B}{\lambda}}$ and  $e^{\frac{-j2\pi R_{z}^B}{\lambda}}$ represent the roundtrip delay corresponding to the $w$th target and $z$th clutter, respectively, where $R_{w}^B$ and $R_{z}^B$ are the corresponding sum distances of the transmitter-$w$th target-receiver pair and transmitter-$z$th clutter-receiver pair, respectively.}
The quantities $\mathbf{c}_w \in \mathbb{C}^{N_t\times 1}$ and $\mathbf{d}_{w} \in \mathbb{C}^{N_r\times 1}$ are the transmit and receive array response vectors, respectively, corresponding to the $w$th target, which are given as
\begin{align}
\mathbf{c}_w =\Big[
1,e^{\frac{-j2\pi d_{t}}{{\lambda}}\text{sin}({\theta}_{w})},\cdots, e^{\frac{-j2\pi d_{t}}{{\lambda}}(N_t-1)\text{sin}({\theta}_{w})}\Big]^T,
\label{tarv} \\
\mathbf{d}_{w} =  \Big[
1,e^{\frac{-j2\pi d_{r}}{{\lambda}}\text{sin}({\phi}_{w})},\cdots, e^{\frac{-j2\pi d_{r}}{{\lambda}}(N_r-1)\text{sin}({\phi}_{w})}\Big]^T,
\label{rarv}
\end{align}
where $d_t, \text{and}\ d_r$ denote the transmit and receive antenna spacings, respectively, whereas $\theta_{w}, \phi_{w}$ denote the AoD and AoA corresponding to the $w$th target, respectively. 
The quantities $\mathbf{c}_z \in \mathbb{C}^{N_t\times 1}$ and $\mathbf{d}_z \in \mathbb{C}^{N_r\times 1}$ are the transmit and receive array response vectors, respectively, corresponding to the $z$th clutter. Finally, the quantity $\mathbf{Q} \in \mathbb{C}^{N_r\times U}$ is the complex additive white Gaussian noise (AWGN) matrix, the elements of which are independently and identically distributed with zero mean and variance $\sigma^2$.
Upon vectorizing the expression in \eqref{recv_node}, the system model can be recast as 
\begin{equation}
     \mathbf{y} = \boldsymbol{\widetilde{\Psi}} \boldsymbol{\tilde{\beta}} + \mathbf{q} , \label{vec_truemodel}
\end{equation}
where $\mathbf{y}= \text{vec}\left(\underline{\mathbf{Y}}\right) \in \mathbb{C}^{N_rU\times 1}$ , $\mathbf{q} = \text{vec}\left(\mathbf{Q}\right)\in\mathbb{C}^{N_rU\times 1}$, $\boldsymbol{\widetilde{\Psi}} =[\boldsymbol{\widetilde{\psi}}_1,\cdots,\boldsymbol{\widetilde{\psi}}_{W'}] \in \mathbb{C}^{N_rU \times W'}$ is the sensing matrix, \textcolor{black}{$\boldsymbol{\tilde{\beta}} =[e^{\frac{-j2\pi R_{1}^B}{\lambda}}\beta_1,\cdots,e^{\frac{-j2\pi R_{W'}^B}{\lambda}}\beta_{W'}] \in \mathbb{C}^{W' \times 1}$ is the RCS coefficient vector} and $W'=W+N_{cl}$. Each column of the sensing matrix is given by $\boldsymbol{\tilde{\psi}}_{w'} =  \text{vec}\left(\mathbf{d}_{w'}\mathbf{c}^T_{w'}\mathbf{X}_{w'}\right)$. 
As observed from Fig. \ref{system_model}, the parameters to be estimated for localization of the $w$th mobile target in the bistatic scenario are AoD $\theta_w$, AoA $\phi_w$ and Doppler shift $\omega_w$. 
\begin{remark}
Note that one can also utilize the classic cyclic prefix-orthogonal frequency division multiplexing (CP-OFDM) waveform for target localization \cite{9166743, 10736660} in bistatic mmWave MIMO ISAC systems in the presence of clutter. The $l$th received sample at the $r$th receive antenna corresponding to the $m$th OFDM symbol can then be expressed as
\begin{align}
    {y}_{m}^{r}[l] &= \sum_{w=1}^{W} \beta_{w} \left[\mathbf{d}_w\right]_r  e^{j2\pi  f_w mT_{s}}e^{j2\pi T_s \frac{l}{N_c} f_w }   \mathbf{c}^T_w \times \nonumber \\
 & \frac{1}{\sqrt{N_c}} \sum_{n=0}^{N_c-1} \mathbf{x}_{n,m} e^{j2\pi n \frac{l}{N_c}} e^{-j2\pi n \Delta f \tau_w} +\nonumber \\
  & \hspace{-10pt}\sum_{z=1}^{N_{cl}} \beta_{z} \left[\mathbf{d}_z\right]_r   \mathbf{c}^T_z 
  \frac{1}{\sqrt{N_c}} \sum_{n=0}^{N_c-1} \mathbf{x}_{n,m} e^{j2\pi n \frac{l}{N_c}} e^{-j2\pi n \Delta f \tau_z} ,\label{y_l_r}
\end{align}
where $\mathbf{x}_{n,m} \in \mathbb{C}^{N_t \times 1}$ is the complex transmit data vector corresponding to the $n$th subcarrier of the $m$th OFDM symbol, $T$ is the total OFDM symbol duration given by $T = T_s+ T_{cp}, T_s$ is the OFDM symbol duration, $T_{cp}$ is the cyclic prefix duration and $\Delta f= \frac{1}{T_s}$ is the subcarrier spacing. The spectral steering vector and inter-carrier interference (ICI) phase rotation vector are defined as follows:
\begin{align}
{\mathbf{b}}_{w'} &= \left[1, e^{-j2\pi \Delta f \tau_{w'}},\cdots,e^{-j2\pi (N_c-1) \Delta f \tau_{w'}}\right]^T  \in \mathbb{C}^{N_c \times 1}, \nonumber \\
{\mathbf{a}}_{w'} &=\left[ 1, e^{j2\pi T_s \frac{1}{N_c} f_{w'} }, \cdots, e^{j2\pi T_s \frac{N_c-1}{N_c} f_{w'}} \right]^T \in \mathbb{C}^{N_c \times 1}.
\end{align}
As shown in our previous work \cite{10683028}, upon stacking the received samples $y_m^r[l], 1 \leq l \leq N_c,$ the received vector at the receiver in the $m$th OFDM symbol is given as
\begin{align}
   \mathbf{y}_m &= \sum_{w'=1}^{W'} \beta_{w'} \bigg\{ {\mathbf{d}}_{w'} \otimes \bigg\{{\mathbf{a}}_{w'}\odot\bigg(\mathbf{F}_{N_c}^H\big({\mathbf{b}}_{w'}\odot \big({\mathbf{X}}_m {\mathbf{c}}_{w'}\big)\big)\bigg)\bigg\}\bigg\} \nonumber  \\
&+ \mathbf{q}_m =\widetilde{\boldsymbol{\Psi}}\widetilde{\boldsymbol{\beta}}+\mathbf{q}_m\in \mathbb{C}^{{N_cN_r} \times 1},\label{ymt}
\end{align}
where each column of the sensing matrix $\widetilde{\boldsymbol{\Psi}}$ is given as $\widetilde{\boldsymbol{\psi}}_{w'} = \bigg\{ {\mathbf{d}}_{w'} \otimes \bigg\{{\mathbf{a}}_{w'}\odot\bigg(\mathbf{F}_{N_c}^H\big({\mathbf{b}}_{w'}\odot \big({\mathbf{X}}_m {\mathbf{c}}_{w'}\big)\big)\bigg)\bigg\}\bigg\}$and $\tilde{\boldsymbol{\beta}} = \left[\beta_1,\beta_2,\cdots,\beta_{W'} \right] \in \mathbb{C}^{W' \times 1}$ is the RCS coefficient vector. It can be readily inferred that \eqref{ymt} is similar to \eqref{vec_truemodel}.
\end{remark}
\subsection{\textcolor{black}{Sparse Problem Formulation}}
To formulate the joint AoA, AoD and Doppler shift estimation as a sparse problem, let the imaging scene be divided into $G$ AoA bins, $G$ AoD bins and $G$ Doppler bins. The uniform sampling of the 3D grid leads to $G^3$ $3$-tuples: $\big\{(\theta_1,\phi_1,\omega_1), (\theta_1,\phi_1,\omega_2),\cdots,(\theta_1,\phi_1,\omega_G),(\theta_1,\phi_2,\omega_1),\cdots,$ $(\theta_G,\phi_G,\omega_G)\big\}$. However, this leads to high computational complexity for the ensuing search and a high correlation between the adjacent grid points \cite{8880558}. For example, two nearby $3$-tuples $(30^{\circ} , 50^{\circ}, 0.075), (25^{\circ} , 50^{\circ}, 0.25)$ share the same AoA leading to a  high correlation between these two bases of the dictionary matrix. This significantly degrades the angle and Doppler shift estimation accuracies. This motivates us to perform non-uniform sampling of the $3$D grid, where no two grid points share the same AoA, AoD and Doppler shift, i.e. we have
\begin{align}
    \theta_{g_i} \neq \theta_{g_j},  \phi_{g_i} \neq \phi_{g_j},  \omega_{g_i} \neq \omega_{g_j}, \forall g_i \neq g_j.
\end{align}
Therefore, the received signal in \eqref{vec_truemodel} can be expressed as 
\begin{equation}
\mathbf{ {y}}= \mathbf{\Psi}\boldsymbol{\beta} + \mathbf{ {q}}
\label{recv_bin_vec},
\end{equation}
where $\mathbf{\Psi} = \Big[\boldsymbol{\psi}_1, \boldsymbol{\psi}_2, \cdots,\boldsymbol{\psi}_g, \cdots, \boldsymbol{\psi}_{G}\Big] \in \mathbb{C}^{N_rU\times G}$ is the dictionary matrix. Each column vector $\boldsymbol{\psi}_g \in \mathbb{C}^{N_rU\times 1}$ of the dictionary matrix $\boldsymbol{\Psi}$ is expressed as 
\begin{equation}
\boldsymbol{\psi}_g = \text{vec}\left(\mathbf{d}_g\mathbf{c}^T_g\mathbf{X}_g\right).
\label{eqn:13}
\end{equation}
Furthermore, $\boldsymbol{\beta} = 
\Big[
\beta_1,\beta_2, \cdots,\beta_{G}
\Big]^T\in \mathbb{C}^{G \times1}$ is the vector of the complex RCS coefficients.  
Note that, the vector $\boldsymbol{\beta}$ is sparse in nature since it has a support of only $W'$ non-zero elements within its support length $G$, with $G \gg W'$. 
 The target parameters ${\boldsymbol{\eta}} = \left[\boldsymbol{\theta}, \boldsymbol{\phi}, \boldsymbol{\omega}\right] \in \mathbb{R}^{3W' \times 1}$ typically do not lie on the grid of the dictionary matrix $\boldsymbol{\Psi}$ due to the non-uniform sampling of the $3D$ grid.
 This grid mismatch often leads to large estimation errors, thus, degrading the performance of the system.  This off-grid gap can be handled by linearly approximating the steering vectors using a first-order Taylor expansion as follows
\begin{align}
   \boldsymbol{\psi}  \left( \theta_{w'}, \phi_{w'}, \omega_{w'} \right) &= \boldsymbol{\psi} \left( \tilde{\theta}_{g}, \tilde{\phi}_{g}, \tilde{\omega}_{g} \right)  +  \boldsymbol{\xi}_t \left( \tilde{\theta}_{g}, \tilde{\phi}_{g}, \tilde{\omega}_{g} \right) \times \nonumber \\
    & \hspace{-20pt} \left( \theta_{w'} - \tilde{\theta}_{g}\right)  + \boldsymbol{\xi}_r \left( \tilde{\theta}_{g}, \tilde{\phi}_{g}, \tilde{\omega}_{g} \right) \left( \phi_{w'} - \tilde{\phi}_{g}\right)  \nonumber \\
    & \hspace{-20pt} +\boldsymbol{\xi}_{\omega} \left( \tilde{\theta}_{g}, \tilde{\phi}_{g}, \tilde{\omega}_{g} \right) \left( \omega_{w'} - \tilde{\omega}_{g}\right),
\end{align}
where we have:
\begin{align}
    \boldsymbol{\xi}_t  \left( \tilde{\theta}_{g}, \tilde{\phi}_{g}, \tilde{\omega}_{g} \right) &= {\frac{\partial \boldsymbol{\psi}  \left( \theta_{w'}, \tilde{\phi}_{g}, \tilde{\omega}_{g} \right) }{\partial  \theta_{w'}}} \bigg | _{\theta_{w'} = \tilde{\theta}_{g}} , \\
    \boldsymbol{\xi}_r  \left( \tilde{\theta}_{g}, \tilde{\phi}_{g}, \tilde{\omega}_{g}\right) &= {\frac{\partial \boldsymbol{\psi}  \left( \tilde{\theta}_{g}, \phi_{w'}, \tilde{\omega}_{g} \right) }{\partial  \phi_{w'}}} \bigg | _{\phi_{w'} = \tilde{\phi}_{g}} , \\
    \boldsymbol{\xi}_{\omega} \left( \tilde{\theta}_{g}, \tilde{\phi}_{g}, \tilde{\omega}_{g} \right) &= {\frac{\partial \boldsymbol{\psi}  \left( \tilde{\theta}_{g}, \tilde{\phi}_{g}, {\omega}_{w'} \right) }{\partial  \omega_{w'}}} \bigg | _{\omega_{w'} = \tilde{\omega}_{g}}.
\end{align}
Here, $\tilde{\theta}_g, \tilde{\phi}_g, \tilde{\omega}_g$ are the nearest grid points to the $w$th target $\left( \theta_w, \phi_w, \omega_w \right)$. Upon assuming $\boldsymbol{\Xi}_t  = \left[ \boldsymbol{\xi}_t  \left( \tilde{\theta}_{1}, \tilde{\phi}_{1}, \tilde{\omega}_{1} \right), \cdots, \boldsymbol{\xi}_t  \left( \tilde{\theta}_{G}, \tilde{\phi}_{G}, \tilde{\omega}_{G} \right) \right],\ \boldsymbol{\Xi}_r  = \left[ \boldsymbol{\xi}_r  \left( \tilde{\theta}_{1}, \tilde{\phi}_{1}, \tilde{\omega}_{1} \right), \cdots, \boldsymbol{\xi}_r  \left( \tilde{\theta}_{G}, \tilde{\phi}_{G}, \tilde{\omega}_{G} \right) \right],\ \boldsymbol{\Xi}_{\omega}  = \left[ \boldsymbol{\xi}_{\omega}  \left( \tilde{\theta}_{1}, \tilde{\phi}_{1}, \tilde{\omega}_{1} \right), \cdots, \boldsymbol{\xi}_{\omega}  \left( \tilde{\theta}_{G}, \tilde{\phi}_{G}, \tilde{\omega}_{G} \right) \right]$, the off-grid model for ${\mathbf{y}}$ is formulated as 
\begin{align}
    \mathbf{y} &= \left(\boldsymbol{\Psi}(\tilde{\boldsymbol{\theta}}, \tilde{\boldsymbol{\phi}}, \tilde{\boldsymbol{\omega}}) + \boldsymbol{\Xi}_t \text{diag}(\boldsymbol{\epsilon}_t) + \boldsymbol{\Xi}_r \text{diag}(\boldsymbol{\epsilon}_r) + \boldsymbol{\Xi}_{\omega}\text{diag}(\boldsymbol{\epsilon}_{\omega})\right)\boldsymbol{\beta} \nonumber \\
    & + \mathbf{q} = \mathbf{D}\left(\boldsymbol{\epsilon}_t,\boldsymbol{\epsilon}_r,\boldsymbol{\epsilon}_{\omega}\right)\boldsymbol{\beta} + \mathbf{q} ,
\end{align}
where 
\begin{align} \boldsymbol{\epsilon}_t  &=\begin{cases}\theta _{w}- \tilde \theta _{g}, &w=1,2,\ldots, W' \\ 0, & \mathrm {otherwise}\end{cases},\\
\boldsymbol{\epsilon}_r  &=\begin{cases}\phi _{w}- \tilde \phi _{g}, &w=1,2,\ldots, W' \\ 0, & \mathrm {otherwise}\end{cases}, \\
 \boldsymbol{\epsilon}_{\omega}  &=\begin{cases}\omega _{w}- \tilde \omega _{g}, &w=1,2,\ldots, W' \\ 0, & \mathrm {otherwise}\end{cases}.
\end{align}
Here $\boldsymbol{\epsilon}$ denotes the gap between the actual parameter value and its closest grid point. 
Next, we formulate a row sparse estimation model for the mmWave MIMO radar considered with the aid of multiple transmission snapshots \cite{8529199}. The received vector corresponding to the $l$th snapshot, $1 \leq l \leq L$, where $L$ denotes the total number of snapshots, is expressed as
\begin{equation}
\mathbf{y}_{l} = \mathbf{D}\boldsymbol{\beta}_{l} + \mathbf{q}_{l}.
\label{eqn:15}
\end{equation} 
Upon concatenating the received vector over $L$ snapshots, so that $\mathbf{{Y}} = \Big[
\mathbf{ {y}}_{1},\mathbf{ {y}}_{2},\ldots \mathbf{ {y}}_{L}
\Big] \in \mathbb{C}^{N_rU\times L}$, the output signal matrix is given by
\begin{equation}
\mathbf{{Y}}= \mathbf{D} \mathbf{B}+ \mathbf{{Q}}
\label{mmv_prb}.
\end{equation}
The targets are assumed to be stationary with respect to the tuple for the duration $T_c=LUT_p$, where $T_p$ is the sub-pulse interval. Under this assumption, it can be observed that the vectors $\boldsymbol{\beta}_{l}$ have a common sparsity profile, which in turn makes the matrix  $\mathbf{B} = \Big[
\boldsymbol{\beta}_{1}, \boldsymbol{\beta}_{2}, \cdots, \boldsymbol{\beta}_{L}
\Big] \in \mathbb{C}^{G \times L}$ row-sparse in nature. The
 next section describes the SBL algorithm conceived for our
 joint angle-velocity-RCS estimation problem.
\section{Bayesian Inference for Joint Target parameter estimation and Localization} \label{BL}
The powerful SBL scheme is employed for
exploiting the 3D sparsity and for the estimation of the sparse RCS matrix ${\mathbf{B}}$.
The SBL framework imposes a parameterized Gaussian prior on each $g$th row of the unknown RCS sparse matrix ${\mathbf{B}}$, denoted by $\boldsymbol{\beta}_g$, as follows:
\begin{align}
    p(\boldsymbol{\beta}_g;z_g) &\sim \mathcal{CN}\left(\mathbf{0}_{1\times L},z_g^{-1}\mathbf{I}_L\right) \nonumber \\
    &= \prod_{l=1}^{L} \frac{z_g}{\pi} \text{exp}\left(-z_g \vert\beta_{g,l}\vert^2\right),
    \label{prior_SBL}
\end{align}
where the hyperparameter $z_g, 0\leq g \leq G-1,$ specifies the precision matrix of the multivariate prior associated with $\boldsymbol{\beta}_g$. The prior for the matrix $\mathbf{{B}}$ is given by: 
\begin{equation}
    p(\mathbf{{B}};\mathbf{z}) = \prod_{g=1}^{G} p(\boldsymbol{\beta}_g;z_g). \label{prior_MSBL}
\end{equation}
Observe from \eqref{prior_SBL} that as the hyperparameter obeys $z_g \rightarrow \infty$, the associated $g$th row of the RCS coefficient matrix follows $\boldsymbol{\beta}_g \rightarrow \mathbf{0}_{L \times 1}$ \cite{1315936}. Therefore, the estimation of the hyperparameter matrix is crucial for the estimation of the sparse matrix $\mathbf{B}$.
However, the received signal model is over-parameterized, since the number of observations is close to the total number of parameters to be estimated. For
this purpose, the hyperparameter vector $\mathbf{z} = (z_0,...,z_{G-1})^T $ is
constrained by imposing
a Gamma prior, that is,
\begin{align}
    p(\mathbf{z})&=\prod_{g=0}^{G-1} \text{Gamma}(z_g\vert a+1,b) = \prod_{g=0}^{G-1} \frac{b^a{z}_g^{a}e^{-b{z_g}}}{\Gamma(a+1)},
\end{align}
where
the Gamma function obeys $\Gamma(a)=\int_0^{\infty}t^{a-1}e^{-t}dt$.
To render these priors non-informative (i.e. flat), we assign these parameters very small values, e.g. $a=b=10^{-4}$, and assume that $\boldsymbol{\epsilon}_t, \boldsymbol{\epsilon}_r, \boldsymbol{\epsilon}_{\omega}$ have non-informative uniform priors. One can now maximize the posterior distribution of the hidden variables $\mathbf{\Omega}_h =\{\mathbf{B},\mathbf{z}, \boldsymbol{\epsilon}_t, \boldsymbol{\epsilon}_r, \boldsymbol{\epsilon}_{\omega}\}$ conditioned on the observed data $\mathbf{Y}$ as follows
\begin{align}
    \mathbf{\Omega}_h =  \text{arg}\ \underset{\mathbf{\Omega}_h}{\text{max}}\ \text{ln}\ p\left ( \mathrm{\boldsymbol \Omega }_h |  \mathbf {Y}\right) \equiv   \text{arg}\ \underset{\mathbf{\Omega}_h}{\text{max}}\ \text{ln}\ p\left ({{ \mathbf {Y},\mathrm{\boldsymbol \Omega }_h } }\right).
\end{align}
Bayesian inference requires the computation of the posterior distribution 
 \begin{align}
      p\left(\mathbf{\Omega}_h\vert\mathbf{Y}\right)&=\frac {p\left ({{ \mathbf {Y},{\mathrm{\boldsymbol \Omega }_h}} }\right)}{\int {p\left ({{ \mathbf {Y},{\mathrm{\boldsymbol \Omega }_h}} }\right)d{\mathrm{\boldsymbol \Omega }_h}}} \nonumber\\  &\hspace{-40pt}=\frac{p\left(\mathbf{Y}\vert\mathbf{z},\boldsymbol{\epsilon}_t, \boldsymbol{\epsilon}_r, \boldsymbol{\epsilon}_{\omega}\right)p(\mathbf{z})}{p\left(\mathbf{Y}\right)}.\label{condPost}
 \end{align}
Here, the numerator denotes the joint probability distribution, while the denominator, termed marginal likelihood or Bayesian evidence, serves as the normalization constant.
We further define the likelihood of the observation as
\begin{equation} p({\mathbf{Y}}| \mathbf{z}, \boldsymbol{\epsilon}_t, \boldsymbol{\epsilon}_r, \boldsymbol{\epsilon}_{\omega}) = {\mathcal C}{\mathcal {N}}({\mathbf{Y}}|\mathbf{DB},{{\sigma}^2}).\end{equation}
However, since the multidimensional integration in \eqref{condPost} cannot be carried out analytically, a closed-form expression of the posterior  $p\left(\mathbf{\Omega}_h\vert\mathbf{Y}\right)$ is challenging to obtain. To circumvent this issue, we harness the block MM algorithm of \cite{7547360}. A surrogate (lower bound) function is computed for the objective function $\ln p\left ({{ \mathbf {Y}, \mathrm{\boldsymbol \Omega }_h } }\right)$, which is maximized with respect to the parameters $\mathbf{z},  \boldsymbol{\epsilon}_t, \boldsymbol{\epsilon}_r, \boldsymbol{\epsilon}_{\omega}$. The surrogate function of $\text{ln}\ p\left(\mathbf{Y},\mathbf{z}, \boldsymbol{\epsilon}_t, \boldsymbol{\epsilon}_r, \boldsymbol{\epsilon}_{\omega}\right)$, at a fixed point $\left(\dot{\mathbf{z}},\dot{\boldsymbol{\epsilon}}_t, \dot{\boldsymbol{\epsilon}}_r, \dot{\boldsymbol{\epsilon}}_{\omega}\right)$, is given as 
\begin{align}
    &\mathcal{L}\left(\mathbf{z}, \boldsymbol{\epsilon}_t, \boldsymbol{\epsilon}_r, \boldsymbol{\epsilon}_{\omega} \vert \dot{\mathbf{z}},\dot{\boldsymbol{\epsilon}}_t, \dot{\boldsymbol{\epsilon}}_r, \dot{\boldsymbol{\epsilon}}_{\omega} \right) = \nonumber \\
    &\int {p\left(\mathbf{B}\vert \mathbf{Y},  \dot{\mathbf{z}}, \dot{\boldsymbol{\epsilon}}_t, \dot{\boldsymbol{\epsilon}}_r, \dot{\boldsymbol{\epsilon}}_{\omega}\right)} \ln \frac {p\left (\mathbf{B,Y}, \mathbf{z}, \boldsymbol{\epsilon}_t, \boldsymbol{\epsilon}_r, \boldsymbol{\epsilon}_{\omega}\right)} {p\left(\mathbf{B}\vert \mathbf{Y},  \dot{\mathbf{z}},\dot{\boldsymbol{\epsilon}}_t, \dot{\boldsymbol{\epsilon}}_r, \dot{\boldsymbol{\epsilon}}_{\omega}\right)} d{\mathbf{B}}.
\end{align}
The parameters $\mathbf{z}, \boldsymbol{\epsilon}_t, \boldsymbol{\epsilon}_r, \boldsymbol{\epsilon}_{\omega}$ are updated as 
\begin{align}
    \mathbf{z}^{(k+1)} &=\text{arg}\underset{\mathbf{z}}{\text{max}}\  \mathcal{L}\bigg(\mathbf{z}, \boldsymbol{\epsilon}_t^{(k)}, \boldsymbol{\epsilon}_r^{(k)}, \boldsymbol{\epsilon}_{\omega}^{(k)} \vert \mathbf{z}^{(k)}, \boldsymbol{\epsilon}_t^{(k)}, \boldsymbol{\epsilon}_r^{(k)}, \boldsymbol{\epsilon}_{\omega}^{(k)} \bigg), \label{z_update}\\
     \boldsymbol{\epsilon}_t^{(k+1)} &=\text{arg}\underset{\boldsymbol{\epsilon}_t}{\text{max}}\  \mathcal{L}\bigg(\mathbf{z}^{(k+1)}, \boldsymbol{\epsilon}_t, \boldsymbol{\epsilon}_r^{(k)}, \boldsymbol{\epsilon}_{\omega}^{(k)} \nonumber \\
     & \vert \mathbf{z}^{(k+1)},\boldsymbol{\epsilon}_t^{(k)}, \boldsymbol{\epsilon}_r^{(k)},  \boldsymbol{\epsilon}_{\omega}^{(k)} \bigg), \label{et_update}\\ 
      \boldsymbol{\epsilon}_r^{(k+1)} &=\text{arg}\underset{\boldsymbol{\epsilon}_r}{\text{max}}\  \mathcal{L}\bigg(\mathbf{z}^{(k+1)},\boldsymbol{\epsilon}_t^{(k+1)}, \boldsymbol{\epsilon}_r, \boldsymbol{\epsilon}_{\omega}^{(k)} \nonumber \\ &\vert \mathbf{z}^{(k+1)}, \boldsymbol{\epsilon}_t^{(k+1)}, \boldsymbol{\epsilon}_r^{(k)}, \boldsymbol{\epsilon}_{\omega}^{(k)} \bigg), \label{er_update}\\ 
      \boldsymbol{\epsilon}_{\omega}^{(k+1)} &=\text{arg}\underset{\boldsymbol{\epsilon}_{\omega}}{\text{max}}\  \mathcal{L}\bigg(\mathbf{z}^{(k+1)}, \boldsymbol{\epsilon}_t^{(k+1)}, \boldsymbol{\epsilon}_r^{(k+1)}, \nonumber \\ &  \boldsymbol{\epsilon}_{\omega} \vert \mathbf{z}^{(k+1)}, \boldsymbol{\epsilon}_t^{(k+1)}, \boldsymbol{\epsilon}_r^{(k+1)}, \boldsymbol{\epsilon}_{\omega}^{(k)} \bigg).  \label{ew_update}
\end{align}
The posterior distribution of each column of $\mathbf{B}$ is given by $q (\boldsymbol{\beta}_l) \sim \mathcal{CN}\left( \boldsymbol{\mu}_{l}^{(k+1)}, \boldsymbol{\Sigma}^{(k+1)} \right)$, where we have
\begin{align}
    \boldsymbol{\Sigma}^{(k+1)}&=\left(\frac{1}{\sigma^2}\mathbf{D}^H\mathbf{D}+\mathbf{Z}^{(k+1)}\right)^{-1},\\     \mathbf{U}^{(k+1)}&=\left[ \boldsymbol{\mu}_{1}^{(k+1)},\ldots, \boldsymbol{\mu}_{L}^{(k+1)} \right] =\frac{1}{\sigma^2}\boldsymbol{\Sigma}^{(k+1)}\mathbf{D}^H\mathbf{Y} \label{mu_update}.
\end{align}
 \begin{figure}[h]
\centering \includegraphics[scale=0.6]{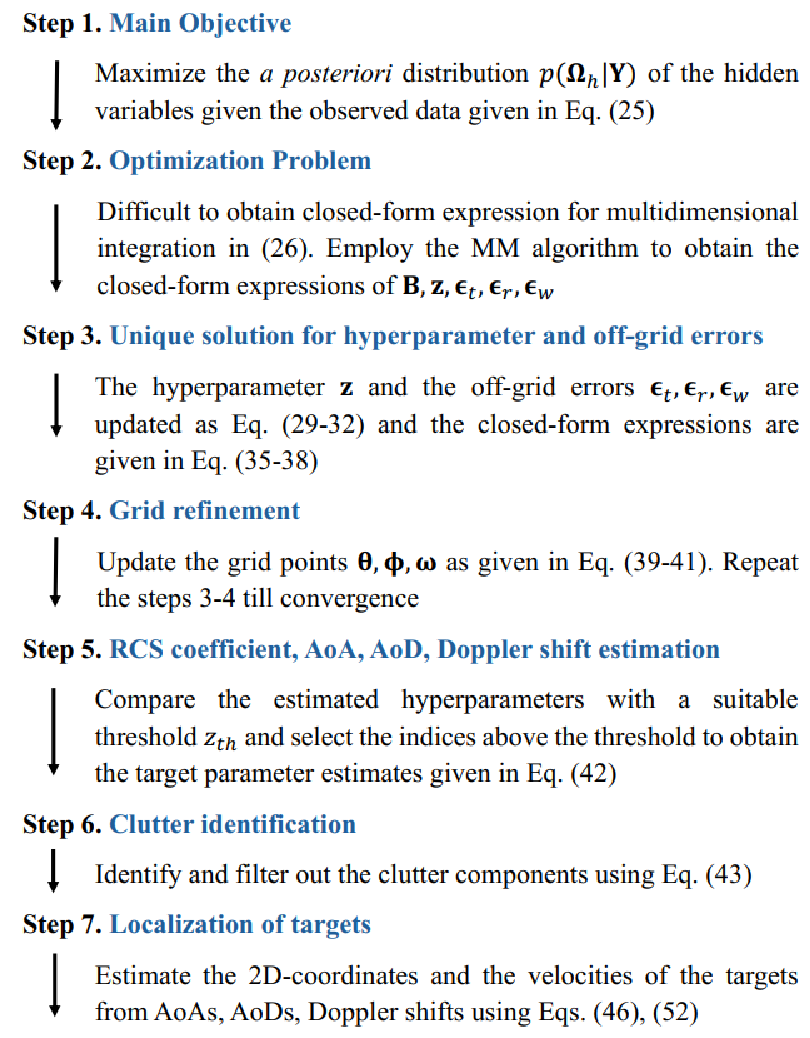}
\caption{\textcolor{black}{Flow of the proposed algorithm}}
\label{flowchart}
\end{figure}
The update expressions of the parameters $\mathbf{z},  \boldsymbol{\epsilon}_t, \boldsymbol{\epsilon}_r, \boldsymbol{\epsilon}_{\omega}$ are given as follows:
\begin{align}
        z_g^{(k+1)} &= \frac{a+L}{b + \sum_{l=1}^L \Upsilon_l \left( \boldsymbol{\epsilon}_t^{(k)}, \boldsymbol{\epsilon}_r^{(k)}, \boldsymbol{\epsilon}_{\omega}^{(k)}\right)}, \label{z_update_final} \\
      \boldsymbol{\epsilon}_t ^{(k+1)} &= \left( \Re\left\{ \boldsymbol{\Xi}_t^H\boldsymbol{\Xi}_t \odot (\mathbf{U}\mathbf{U}^H + L\boldsymbol{\Sigma})  \right\} \right)^{\dagger} \mathbf{p}_t, \label{et_update_final}    \\
     \boldsymbol{\epsilon}_r ^{(k+1)} &= \left( \Re\left\{ \boldsymbol{\Xi}_r^H \boldsymbol{\Xi}_r \odot (\mathbf{U}\mathbf{U}^H + L\boldsymbol{\Sigma})  \right\} \right)^{\dagger} \mathbf{p}_r, \label{er_update_final} \\
     \boldsymbol{\epsilon}_{\omega} ^{(k+1)} &= \left( \Re\left\{ \boldsymbol{\Xi}_{\omega}^H \boldsymbol{\Xi}_{\omega} \odot (\mathbf{U}\mathbf{U}^H + L\boldsymbol{\Sigma})  \right\} \right)^{\dagger} \mathbf{p}_{\omega}, \label{ew_update_final}
\end{align}
where  $\Upsilon_l \left( \boldsymbol{\epsilon}_t^{(k)}, \boldsymbol{\epsilon}_r^{(k)}, \boldsymbol{\epsilon}_{\omega}^{(k)}\right) = \boldsymbol{\mu}_l \boldsymbol{\mu}^H_l + \boldsymbol{\Sigma}$, $\mathbf{p}_t = \Re\left\{\sum_{l=1}^L \text{diag}(\boldsymbol{\mu}_l)\boldsymbol{\Xi}_t^H\left( \mathbf{y}_l - \boldsymbol{\Psi}_t \boldsymbol{\mu}_l\right)  \right\} - L \Re\left\{ \text{diag} \left( \boldsymbol{\Xi}_t^H \boldsymbol{\Psi}_t  \boldsymbol{\Sigma}\right) \right\}$ with $\boldsymbol{\Psi}_{t} =  \boldsymbol{\Psi}(\tilde{\boldsymbol{\theta}}, \tilde{\boldsymbol{\phi}}, \tilde{\boldsymbol{\omega}}) + \boldsymbol{\Xi}_r \text{diag}(\boldsymbol{\epsilon}_r) + \boldsymbol{\Xi}_{\omega}\text{diag}(\boldsymbol{\epsilon}_{\omega})$, $\mathbf{p}_r = \Re\left\{\sum_{l=1}^L \text{diag}(\boldsymbol{\mu}_l)\boldsymbol{\Xi}_r^H\left( {\mathbf{y}}_l - \boldsymbol{\Psi}_r \boldsymbol{\mu}_l\right)  \right\} - L \Re\left\{ \text{diag} \left( \boldsymbol{\Xi}_r^H \boldsymbol{\Psi}_r \boldsymbol{\Sigma}\right) \right\}$ with $\boldsymbol{\Psi}_{r} =  \boldsymbol{\Psi}(\tilde{\boldsymbol{\theta}}, \tilde{\boldsymbol{\phi}}, \tilde{\boldsymbol{\omega}})$ + $\boldsymbol{\Xi}_t \text{diag}(\boldsymbol{\epsilon}_t) + \boldsymbol{\Xi}_{\omega}\text{diag}(\boldsymbol{\epsilon}_{\omega})$, 
 $\mathbf{p}_{\omega} = \Re\left\{\sum_{l=1}^L \text{diag}(\boldsymbol{\mu}_l)\boldsymbol{\Xi}_{\omega}^H \left( {\mathbf{y}}_l - \boldsymbol{\Psi}_{\omega} \boldsymbol{\mu}_l\right)  \right\} - L \Re\left\{ \text{diag} \left( \boldsymbol{\Xi}_{\omega}^H \boldsymbol{\Psi}_{\omega} \boldsymbol{\Sigma}\right) \right\}$ with $\boldsymbol{\Psi}_{\omega} =  \boldsymbol{\Psi}(\tilde{\boldsymbol{\theta}}, \tilde{\boldsymbol{\phi}}, \tilde{\boldsymbol{\omega}}) + \boldsymbol{\Xi}_t \text{diag}(\boldsymbol{\epsilon}_t) + \boldsymbol{\Xi}_r\text{diag}(\boldsymbol{\epsilon}_r)$. The detailed derivations are presented in Appendix \ref{proof}.
The grid points are then updated as follows
\begin{align}
    \tilde{\boldsymbol{\theta}}^{(k+1)} =     \tilde{\boldsymbol{\theta}}^{(k)} + \boldsymbol{\epsilon}_t,  \label{theta_update}\\
        \tilde{\boldsymbol{\phi}}^{(k+1)} =     \tilde{\boldsymbol{\phi}}^{(k)} + \boldsymbol{\epsilon}_r, \label{phi_update} \\
     \tilde{\boldsymbol{\omega}}^{(k+1)} =     \tilde{\boldsymbol{\omega}}^{(k)} + \boldsymbol{\epsilon}_{\omega}. \label{omega_update}
\end{align}
The steps described above are repeated until convergence. Since the hyperparameters control the sparsity in $\widehat{\mathbf{B}}$, the performance attained can be further enhanced by thresholding the hyperparameter estimates with respect to a suitably chosen threshold $z_{th}$. The rows $\widehat{\mathbf {B}}(g,:)$ with hyperparameters $z_g$ higher than $z_{th}$ are set to zero. The pruned estimate $\widehat{\mathbf {B}}$ is given by
\begin{align} \widehat{\mathbf {B}}(g,:)= {\begin{cases}\mathbf{U}^{(k+1)}(g,:), & \widehat{z}_g^{(k+1)} < z_{th}\\ \mathbf{0}_{1\times L}, & \text{otherwise}. \end{cases}} \end{align}
Subsequently, instead of updating all the $G$ grid points in  $\eqref{theta_update}, \eqref{phi_update}, \eqref{omega_update}$, one can only update those corresponding to the significant rows of $\widehat{\mathbf{B}}$. Next, we exploit an interesting property to identify the clutter components in the environment. The ground clutter strength decays with the increase in Doppler frequency \cite{zahabi2019one, 6675884}. Thus, the significant clutter components can be assumed to be confined to a maximum Doppler shift of $\omega_{cl}^{*}$,
beyond which the clutter effect can be neglected. In the above formulation, the parameter $\delta$ is set for ensuring that 
\begin{equation}
   \vert\omega_{cl}^{*}\vert \leq \delta \frac{\omega_{\text{max}}}{G},
\end{equation}
where $\omega_{\text{max}}$ is the maximum Doppler shift of any target in the sensing environment. This implies that if the estimated Doppler shift $\widehat{\omega}_{{w'}}^{(k+1)}$ is lower than or equal to the parameter $\delta+1$, i.e. $\widehat{\omega}_{{w'}}^{(k+1)} \leq \delta+1$, then the estimated AoA-AoD-Doppler tuple $\left(\widehat{\theta}_{{w'}}^{(k+1)},\widehat{\phi}_{{w'}}^{(k+1)},\widehat{\omega}_{{w'}}^{(k+1)}\right)$ belongs to the clutter. This step filters out the $N_{cl}$ clutter components from the ${W'}$ signal-plus-clutter components to yield the estimated AoA, AoD and Doppler shift values corresponding to the $W$ targets. Next, we devise an innovative algorithm for localizing the targets by computing their 2D coordinates and velocities from the estimated AoDs, AoAs and Doppler shifts.
\begin{figure}[h]
    \centering
    \includegraphics[width=0.9\linewidth]{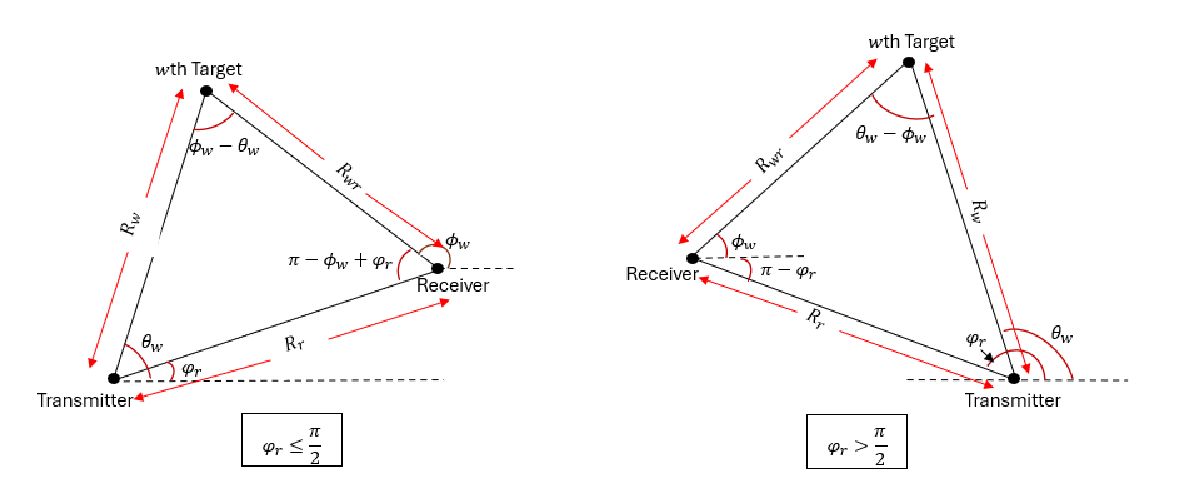}
    \caption{\textcolor{black}{Point representation of transmitter-$w$th target-receiver pair under different conditions of receiver orientation $\varphi_r$}}
    \label{BR}
\end{figure}
\subsection{Localization Algorithm} \label{localization}
\textbf{2D Coordinate estimation}: 
\textcolor{black}{Fig. \ref{BR} represents the point based representation of the transmitter-$w$th target-receiver pair under both conditions of receiver orientation $\varphi_r \leq \frac{\pi}{2}$ and  $\varphi_r > \frac{\pi}{2}$. Using the law of sines for triangles, the distance of the $w$th target from the transmitter $ \widehat{R}_w$ can be formulated as  
\begin{align}
    \widehat{R}_w =
    {\begin{cases}R_r\frac{\text{sin}(\pi - \hat{\phi}_w + \varphi_r)}{\text{sin}( \hat{\phi}_w - \hat{\theta}_w )}, & \varphi_r \leq \frac{\pi}{2}\\  R_r\frac{\text{sin}(\hat{\phi}_w + \pi - \varphi_r)}{\text{sin}( \hat{\theta}_w - \hat{\phi}_w )}, & \varphi_r >\frac{\pi}{2}. \end{cases}}
\end{align}}
Furthermore, since the transmitter is assumed to be located at the origin, the orientation of the $w$th target ${\varphi}_w$ equals the estimated AoD $\hat{\theta}_w $
\begin{equation}
    \widehat{\varphi}_w = \hat{\theta}_w.
\end{equation}
Subsequently, the $x$ and $y$ coordinates of the $w$th target are given respectively as
\begin{align}
    \hat{x}_w = \widehat{R}_w\text{cos}(\widehat{\varphi}_w),   \hat{y}_w = \widehat{R}_w\text{sin}(\widehat{\varphi}_w) \label{locxy}.
\end{align}
\textcolor{black}{\textbf{Sum distance Estimation}: 
Using geometry, the following expressions can be deduced for $\varphi_r \leq \frac{\pi}{2}$
\begin{align}
{R}_{w}\text{cos}\left({\theta}_{w}-\varphi_r\right) + {R}_{wr}\text{cos}\left(\pi - \phi_w + \varphi_r\right) &= R_{r}  \nonumber \\ 
-{R}_{w}\text{sin}\left({\theta}_{w}-\varphi_r\right) +{R}_{wr}\text{sin}\left(\pi - \phi_w + \varphi_r\right) &= 0.
\end{align}
The unknown vector $\mathbf{r} \triangleq \left[{R}_{w},{R}_{wr}\right]^T$ can be determined by solving the pair of linear equations above, and it is given as
\begin{equation}
    \widehat{\mathbf{r}} = \mathbf{P}^{-1}\mathbf{r}_d,
\end{equation}
where 
\begin{align*}
    \mathbf{P}=\begin{bmatrix}
    \text{cos}\left({\theta}_{w}-\varphi_r\right)& \text{cos}\left(\pi - \phi_w + \varphi_r\right)\\
    -\text{sin}\left({\theta}_{w}-\varphi_r\right)& \text{sin}\left(\pi - \phi_w + \varphi_r\right)\\
    \end{bmatrix} , \mathbf{r}_d = \begin{bmatrix}
    {R}_{r} \\0
        \end{bmatrix}.
\end{align*}
The estimated range for the transmitter- $w$th target-receiver pair is thus given as 
\begin{equation}
\widehat{R}_{wr} = R_{r}\frac{\text{sin}\left( {\theta}_{w}-\varphi_r\right)}{\text{sin}\left(\pi - \phi_w + \theta_w\right)}, \  \widehat{R}_{w} = R_{r}\frac{\text{sin}\left( \pi - \phi_w + \varphi_r\right)}{\text{sin}\left(\pi - \phi_w + \theta_w\right)}.
\end{equation}    
The estimated sum distance $\widehat{R}^B_{w} = \widehat{R}_{wr} + \widehat{R}_{w}$ can now be derived as
\begin{equation}
        \widehat{R}^B_{w} = R_{r}\frac{\left[\text{sin}\left( {\theta}_{w}-\varphi_r\right) + \text{sin}\left(  \pi - \phi_w + \varphi_r\right)\right]}{\text{sin}\left(\pi - \phi_w + \theta_w\right)}.
\end{equation}
Similarly, for  $\varphi_r > \frac{\pi}{2}$, the sum distance $\widehat{R}^B_{w}$ is given as
\begin{equation}
        \widehat{R}^B_{w} = R_{r}\frac{\left[\text{sin}\left( {\varphi_r-\theta}_{w}\right) + \text{sin}\left(  \pi - \varphi_r  + \phi_w \right)\right]}{\text{sin}\left(\pi + \phi_w - \theta_w\right)}.
\end{equation}}
\textbf{Velocity estimation}:
The estimated velocity $\widehat{v}_w$ of the $w$th target is expressed as
\begin{equation}
\widehat{v}_w = \frac{\widehat{\omega}_{w}\lambda}{4 \pi T_p}.
\label{velocity_estimation}
\end{equation}
The flow of the proposed algorithm is summarized in Fig. \ref{flowchart}.
\textcolor{black}{\subsection{Complexity Analysis} \label{complexity}
The computational complexity of the proposed algorithm is as follows:
\begin{itemize}
    \item The complexities of calculating $\mathbf{U}$ and $\boldsymbol{\Sigma}$ are $\mathcal{O} \left( LG^2 \right)$ and $\mathcal{O} \left[ G^2.\text{max}(G,N_rU) \right]$, respectively.
    \item The complexity in updating the hyperparameter vector $\mathbf{z}$ is $\mathcal{O} \left( LG \right)$.
    \item Since $\epsilon_t, \epsilon_r \ \text{and} \  \epsilon_\omega$ are jointly sparse with $\mathbf{B}$, the complexity in updating $\epsilon_t, \epsilon_r$ and $\epsilon_\omega$ can be ignored.
\end{itemize}
Therefore, the overall computational complexity of the proposed design is $\mathcal{O} \big[ NG^2 \times $ $\text{max}(G,N_rU) \big]$, where $N$ is the total number of SBL iterations. 
\begin{table}[h]
\centering
\caption{{Computational Complexities of Various Algorithms}}
    \begin{tabular}{|l|r|}
 \hline
 \textbf{Algorithm} & \textbf{Computational Complexity} \\
 \hline
 Proposed Scheme  &   $\mathcal{O} \big[ NG^2 \times $ $\text{max}(G,N_rU) \big]$ \\
\hline
OMP \cite{5895106} &  $\mathcal{O} (LNN_rUG)$  \\
\hline
MUSIC \cite{9512486} & $\mathcal{O} \big( N_r^3U^3 + G^2N_r^2U^2 \big)$ \\
\hline
    \end{tabular}
     \label{Tab:comp}
\end{table}
The computational complexity of the proposed SBL algorithm is also compared with the existing OMP \cite{5895106} and MUSIC \cite{9512486} schemes and summarized in Table \ref{Tab:comp}. Our method and MUSIC have comparable computational complexities, whereas OMP has a lower computational burden than both SBL and MUSIC. However, it can be observed in Section V that the proposed SBL scheme achieves a significant improvement in estimation performance, especially in the case of low
SCNR.}
\textcolor{black}{\section{Cram\'er Rao Bound for AoD, AoA, velocity Estimation and Bayesian Cram\'er Rao Bound for RCS Coefficient Matrix Estimation} \label{crb}
The CRBs for the AoDs, AoAs, velocity estimation and Bayesian CRB (BCRB) for RCS coefficient estimation of the multiple targets are derived next.}
\subsection{CRBs for joint AoD, AoA, Velocity Estimation}
The unknown channel domain parameter vector can be constructed as $\boldsymbol{\eta}=\left[\boldsymbol{\theta}^T, \boldsymbol{\phi}^T, \boldsymbol{\omega}^T \right]^T$, where we have $\boldsymbol{\theta} = [\theta_{1},\theta_{2},\cdots,\theta_{W}] \in \mathbb{R}^{W \times 1}$, $\boldsymbol{\phi} = [\phi_{1},\phi_{2},\cdots,\phi_{W}] \in \mathbb{R}^{W \times 1}$ and $\boldsymbol{\omega} = [\omega_{1},\omega_{2},\cdots,\omega_{W}]\in \mathbb{R}^{W \times 1}$. The Fisher information matrix (FIM) $\mathbf{J}_{\boldsymbol{\xi}} \in \mathbb{R}^{3W\times 3W}$ satisfies the information inequality of \cite{kay1993fundamentals} 
\begin{equation}
    \mathbb{E}\left\lbrace \left( \hat{\boldsymbol{\eta}} - \boldsymbol{\eta} \right)\left( \hat{\boldsymbol{\eta}} - \boldsymbol{\eta} \right)^T \right\rbrace \succcurlyeq \mathbf{J}_{\boldsymbol{\eta}}^{-1},
\end{equation}
for any unbiased estimator $\hat{\boldsymbol{\eta}}$ of ${\boldsymbol{\eta}}$. The FIM $\mathbf{J}_{\boldsymbol{\eta}}$ can be further written as
\begin{equation}
    \mathbf{J}_{\boldsymbol{\eta}} = \begin{bmatrix}
\mathbf{J}_{\boldsymbol{\theta}} & \mathbf{J}_{\boldsymbol{\theta\phi}} & \mathbf{J}_{\boldsymbol{\theta\omega}}\\
\mathbf{J}_{\boldsymbol{\phi\theta}} & \mathbf{J}_{\boldsymbol{\phi}}  &  \mathbf{J}_{\boldsymbol{\phi\omega}}\\
\mathbf{J}_{\boldsymbol{\omega\theta}} &  \mathbf{J}_{\boldsymbol{\omega\phi}} &\mathbf{J}_{\boldsymbol{\omega}}
\end{bmatrix}.	\label{FIM_overall}
\end{equation}
The log-likelihood function $\mathcal{L}\left(\mathbf{Y};\boldsymbol{\eta}\right) = \text{ln}\ p (\mathbf{Y};\boldsymbol{\eta})$ is given by
\begin{align}  
&\mathcal{L}\left(\mathbf{Y};\boldsymbol{\eta}\right) = c_1 - \sum_{l=1}^L\left(\tilde{\mathbf{y}}_l - \mathbf{D}\boldsymbol{\beta}_l \right)^H \left(\tilde{\mathbf{y}}_l - \mathbf{D}\boldsymbol{\beta}_l \right),
\end{align}
where the constant obeys $c_1 = -N_y \text{log} (\pi ) -\text{log}(\boldsymbol{\Sigma}_c^{-1})$. The diagonal entries of the FIM $\mathbf{J}_{\boldsymbol{\xi}}$ are computed next. For a parameter vector, $\boldsymbol{\zeta}= \boldsymbol{\theta}\hspace{0.15cm} \text{or}\hspace{0.15cm}  \boldsymbol{\phi}\hspace{0.15cm}  \text{or}\hspace{0.15cm}  \boldsymbol{\omega}$, the $(e,f)$th element of the FIM $\mathbf{J}_{\boldsymbol{\zeta}}$ is given by
\begin{align}
\left[\mathbf{J}_{\boldsymbol{\zeta}}\right]_{ef}&=- \mathbb{E}_{\mathbf{Y}}\left\{ \frac {\partial^2 \mathcal {L}\left(\mathbf{Y};\boldsymbol{\eta}\right)}{\partial \zeta_e \partial \zeta_f} \right\}\nonumber\\    
    & \hspace{-20pt} =2 \Re \left\{ \sum_{l=1}^L \left(\frac{\partial \left( \mathbf{D}\boldsymbol{\beta}_l \right)^H}{\partial \zeta_e}\right) \left(\frac{\partial \left( \mathbf{D}\boldsymbol{\beta}_l \right)}{\partial \zeta_f}\right)\right\} \nonumber \\
    & \hspace{-20pt} =2 \Re \left\{ \sum_{l=1}^{L}\left(\beta_{e,l}^*\frac{\partial \boldsymbol{\psi}_e^H}{\partial \zeta_e}\right)\left(\frac{\partial \boldsymbol{\psi}_f}{\partial \zeta_f}\beta_{f,l}\right)\right\}.
    \end{align}
Let us now evaluate the off-diagonal entries of the FIM $\mathbf{J}_{\boldsymbol{\eta}}$.
For $1\leq e,f\leq W$, the $(e,f)$th element of the FIMs $\mathbf{J}_{\boldsymbol{\theta\phi}}$, $\mathbf{J}_{\boldsymbol{\theta\omega}}$ and $\mathbf{J}_{\boldsymbol{\phi\omega}}$ can be written as
\begin{align}
    \left[\mathbf{J}_{\boldsymbol{\theta\phi}}\right]_{ef}&=2 \Re \left\{\sum_{l=1}^L \left(\beta_{e,l}^*\frac{\partial \boldsymbol{\psi}_e^H}{\partial \theta_e}\right)\left(\frac{\partial \boldsymbol{\psi}_f}{\partial \phi_f}\beta_{f,l}\right)\right\}, \\
    \left[\mathbf{J}_{\boldsymbol{\theta\omega}}\right]_{ef}&=2 \Re \left\{\sum_{l=1}^L \left(\beta_{e,l}^*\frac{\partial \boldsymbol{\psi}_e^H}{\partial \theta_e}\right) \left(\frac{\partial \boldsymbol{\psi}_f}{\partial \omega_f}\beta_{f,l}\right)\right\}, \\
        \left[\mathbf{J}_{\boldsymbol{\phi\omega}}\right]_{ef}&=2 \Re \left\{\sum_{l=1}^L \left(\beta_{e,l}^*\frac{\partial \boldsymbol{\psi}_e^H}{\partial \phi_e}\right) \left(\frac{\partial \boldsymbol{\psi}_f}{\partial \omega_f}\beta_{f,l}\right)\right\}.
\end{align}
Furthermore, we have: \begin{equation}    \mathbf{J}_{\boldsymbol{\phi\theta}}=\mathbf{J}_{\boldsymbol{\theta\phi}}^H ,\hspace{1cm} \mathbf{J}_{\boldsymbol{\omega\theta}}=\mathbf{J}_{\boldsymbol{\theta\omega}}^H ,\hspace{1cm}
\mathbf{J}_{\boldsymbol{\omega\phi}}=\mathbf{J}_{\boldsymbol{\phi\omega}}^H. \nonumber    
\end{equation}
 \begin{itemize}
    \item \textit{CRB of $\boldsymbol{\theta}$:} 
Let us introduce the vector $\boldsymbol{\rho}=[\boldsymbol{\phi}^T, \boldsymbol{\omega}^T]^T$. The FIM in \eqref{FIM_overall} can be rewritten as
\begin{equation}
     \mathbf{J}_{\boldsymbol{\eta}} = \begin{bmatrix}
\mathbf{J}_{\boldsymbol{\theta}} & \mathbf{J}_{\boldsymbol{\theta\rho}} \\
\mathbf{J}_{\boldsymbol{\rho\theta}} & \mathbf{J}_{\boldsymbol{\rho}}  \\
\end{bmatrix},
\end{equation}
where we have: 
\begin{equation}
    \mathbf{J}_{\boldsymbol{\theta\rho}} = \begin{bmatrix}
\mathbf{J}_{\boldsymbol{\theta\phi}} & \mathbf{J}_{\boldsymbol{\theta\omega}} \\
\end{bmatrix}= \mathbf{J}_{\boldsymbol{\rho\theta}}^H \hspace{0.2cm},\hspace{0.2cm}
     \mathbf{J}_{\boldsymbol{\rho}} = \begin{bmatrix}
\mathbf{J}_{\boldsymbol{\phi}} & \mathbf{J}_{\boldsymbol{\phi\omega}} \\
\mathbf{J}_{\boldsymbol{\omega\phi}} & \mathbf{J}_{\boldsymbol{\omega}}  \\
\end{bmatrix}.
\end{equation} \label{j_thetaDelta}
Invoking the Woodbury matrix identity for the partitioned matrices, we eventually obtain the CRB of $\boldsymbol{\theta}$ as
\begin{equation}
    \mathbb{E}\left\lbrace\vert\vert \hat{\boldsymbol{\theta}} -\boldsymbol{\theta} \vert \vert ^2\right\rbrace \geq \text{Tr}\left(\left(\mathbf{J}_{\boldsymbol{\theta}} -  \mathbf{J}_{\boldsymbol{\theta\rho}}\mathbf{J}_{\boldsymbol{\rho}}^{-1}\mathbf{J}_{\boldsymbol{\rho\theta}}\right)^{-1}\right), \label{crb_theta}
\end{equation}
where $\mathbf{J}_{\boldsymbol{\rho}}^{-1}$ can be obtained by once again invoking the Woodbury matrix identity for partitioned matrices as follows
\begin{equation}
    \mathbf{J}_{\boldsymbol{\rho}}^{-1} = \begin{bmatrix}
        \mathbf{W} & -\mathbf{W}\mathbf{J}_{\boldsymbol{\phi\omega}}\mathbf{J}_{\boldsymbol{\omega}}^{-1} \\ -\mathbf{J}_{\boldsymbol{\omega}}^{-1}\mathbf{J}_{\boldsymbol{\omega\phi}}\mathbf{W} & \left( \mathbf{J}_{\boldsymbol{\omega}} - \mathbf{J}_{\boldsymbol{\omega\phi}}\mathbf{J}_{\boldsymbol{\phi}}^{-1}\mathbf{J}_{\boldsymbol{\phi\omega}} \right)^{-1}\\
    \end{bmatrix}, \label{J_delta_inv}
\end{equation}
where we have $\mathbf{W} = \left( \mathbf{J}_{\boldsymbol{\phi}} - \mathbf{J}_{\boldsymbol{\phi\omega}}\mathbf{J}_{\boldsymbol{\omega}}^{-1}\mathbf{J}_{\boldsymbol{\omega\phi}} \right)^{-1}$.
\item \textit{CRB of $\boldsymbol{\phi}$:} Similar to \eqref{crb_theta}, the CRB of the delay vector $\boldsymbol{\phi}$ can be obtained as
 \begin{equation}
    \mathbb{E}\left\lbrace\vert\vert \hat{\boldsymbol{\phi}} -\boldsymbol{\phi} \vert \vert ^2\right\rbrace \geq \text{Tr}\left(\left(\mathbf{J}_{\boldsymbol{\phi}} -  \mathbf{J}_{\boldsymbol{\phi\rho}}\mathbf{J}_{\boldsymbol{\rho}}^{-1}\mathbf{J}_{\boldsymbol{\rho\phi}}\right)^{-1}\right), \label{crb_phi}
\end{equation}
where, $\boldsymbol{\rho}=[\boldsymbol{\theta}^T, \boldsymbol{\omega}^T]^T$ and correspondingly, we have
\begin{equation}
    \mathbf{J}_{\boldsymbol{\phi\rho}} = \begin{bmatrix}
\mathbf{J}_{\boldsymbol{\phi\theta}} & \mathbf{J}_{\boldsymbol{\phi\omega}} \\
\end{bmatrix}= \mathbf{J}_{\boldsymbol{\rho\phi}}^H ,\hspace{0.2cm}
     \mathbf{J}_{\boldsymbol{\rho}} = \begin{bmatrix}
\mathbf{J}_{\boldsymbol{\theta}} & \mathbf{J}_{\boldsymbol{\theta\omega}} \\
\mathbf{J}_{\boldsymbol{\omega\theta}} & \mathbf{J}_{\boldsymbol{\omega}}  \\
\end{bmatrix}.\label{j_tauDelta}
\end{equation}

The term $\mathbf{J}_{\boldsymbol{\rho}}^{-1}$ can be derived in a similar fashion as shown in \eqref{J_delta_inv}. 
 \item  \textit{CRB of $\mathbf{v}$:} The structure of the FIM $\mathbf{J}_{\boldsymbol{\eta}}$ for the Doppler shift vector $\boldsymbol{\omega}$ can be obtained similar to \eqref{crb_phi} and it is expressed as
 \begin{equation}
    \mathbb{E}\left\lbrace\vert\vert \hat{\boldsymbol{\omega}} -\boldsymbol{\omega} \vert \vert ^2\right\rbrace \geq \text{Tr}\left(\left(\mathbf{J}_{\boldsymbol{\omega}} -  \mathbf{J}_{\boldsymbol{\omega\rho}}\mathbf{J}_{\boldsymbol{\rho}}^{-1}\mathbf{J}_{\boldsymbol{\rho\omega}}\right)^{-1}\right),
\end{equation}
where $\boldsymbol{\rho}=[\boldsymbol{\theta}^T, \boldsymbol{\phi}^T]^T$.  The terms $\mathbf{J}_{\boldsymbol{\omega\rho}} \hspace{0.2cm} \text{and}\hspace{0.2cm}\mathbf{J}_{\boldsymbol{\rho}}$ are similar to \eqref{j_tauDelta}, while the evaluation of $\mathbf{J}_{\boldsymbol{\rho}}^{-1}$ follows \eqref{J_delta_inv}.\\
For obtaining the CRB of the velocity estimate, one can use \eqref{doppler_shift} for defining the transformation matrix $\mathbf{T}=\frac{\partial \boldsymbol{\omega}^T}{\partial \mathbf{v}} \in \mathbb{R}^{W \times W}$ as
\begin{align}
    \mathbf{T}=\frac{\partial \boldsymbol{\omega}^T}{\partial \mathbf{v}} = \begin{bmatrix}
         \frac{\partial {\omega_1}}{\partial {v_1}} & \frac{\partial {\omega_2}}{\partial {v_1}} & \cdots &      \frac{\partial {\omega_W}}{\partial {v_1}}\\
         \frac{\partial {\omega_1}}{\partial {v_2}}& \frac{\partial {\omega_2}}{\partial {v_2}} & \cdots &      \frac{\partial {\omega_W}}{\partial {v_2}} \\
       \vdots &      \vdots &  \ddots &     \vdots \\
        \frac{\partial {\omega_1}}{\partial {v_W}}&       \frac{\partial {\omega_2}}{\partial {v_W}} &  \cdots &      \frac{\partial {\omega_W}}{\partial {v_W}} \\
    \end{bmatrix} = \frac{4 \pi T_p}{\lambda}\label{Tmat} \mathbf{I}.
\end{align}
Therefore, the FIM of the velocity vector can be determined as 
\begin{equation}
   \mathbf{J}_{\mathbf{v}} = \mathbf{T}\mathbf{J}_{\boldsymbol{\omega}}\mathbf{T}^T. \label{FIM_transform_vel}
\end{equation}
Finally, we have the CRB of the velocity estimate given as
\begin{equation}
    \mathbb{E}\left\lbrace\vert\vert \hat{\mathbf{v}} -\mathbf{v} \vert \vert ^2\right\rbrace \geq \text{Tr}\left(\left(\mathbf{J}_{\mathbf{v}} -  \mathbf{J}_{\mathbf{v}\boldsymbol{\rho}}\mathbf{J}_{\boldsymbol{\rho}}^{-1}\mathbf{J}_{\boldsymbol{\rho}\mathbf{v}}\right)^{-1}\right).
\end{equation}
\end{itemize}
\subsection{Bayesian CRB (BCRB) for RCS Coefficient Matrix}
This section derives the BCRB in the mean squared error (MSE) of estimation of the RCS coefficient matrix $\mathbf{B}$. Upon vectorizing \eqref{mmv_prb}, the received signal $\mathbf{y} = \text{vec}\left(\mathbf{Y}\right) \in \mathbb{C}^{LN_rU \times 1}$ can be expressed as
\begin{align}
   \mathbf{y} = \underbrace{\left( \mathbf{I}_{L} \otimes \mathbf{D} \right)}_{\mathbf{\underline{D}}}\boldsymbol{\beta}+ \mathbf{\underline{q}},
\end{align}
where the effective parameter vector obeys $\boldsymbol{\beta} = \text{vec}\left(\mathbf{B} \right) \in \mathbb{C}^{LG \times 1}$ and similarly $\mathbf{\underline{q}} =\text{vec}\left( \mathbf{{Q}}\right)  \in \mathbb{C}^{LN_rU \times 1}$. The Bayesian Fisher information matrix (BFIM) $\mathbf{J}_B \in \mathbb{C}^{LG \times LG}$ is defined as \cite{kay1993fundamentals}
\begin{align}
    \mathbf {J}_{B}\!=\!\underbrace {-\text {E}_{\left ({\mathbf{y},\boldsymbol{\beta}  }\right)}\left \lbrace{ \frac {\partial ^{2}\mathcal {L}\left ({\mathbf{y}\mid \boldsymbol{\beta} ;\mathbf {Z }}\right)}{\partial \boldsymbol{\beta} \partial \boldsymbol{\beta} ^{H}}\!}\right \rbrace }_{\mathbf {J}_{\mathcal {D}}}\underbrace {-\text {E}_{\boldsymbol{\beta} }\left \lbrace{ \frac {\partial ^{2}\mathcal {L}\left ({\boldsymbol{\beta} ; \mathbf{Z}}\right)}{\partial \boldsymbol{\beta}  \partial \boldsymbol{\beta} ^{H}}}\right \rbrace }_{\mathbf {J}_{\mathcal {P}}},
\end{align}
where the matrices  $\mathbf {J}_{\mathcal {P}}$ and $\mathbf {J}_{\mathcal {D}}$ denote the FIMs with respect to the prior density of the parameter vector $\boldsymbol{\beta} $  and the output vector $\mathbf{y}$, respectively. After neglecting the constant terms, 
\begin{figure}[tbh]
\subfloat[]{\includegraphics[scale=0.318]{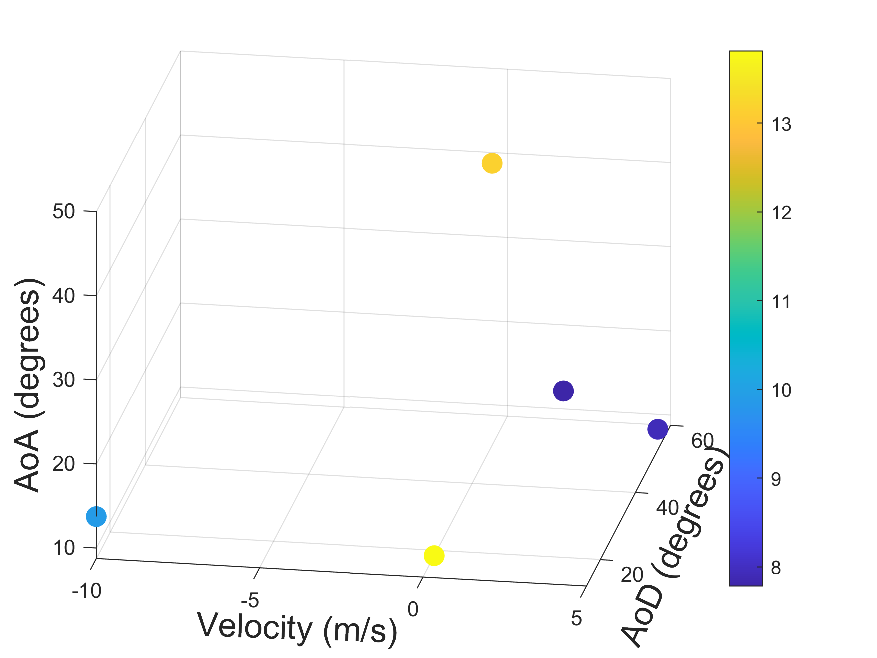}}
\subfloat[]{\includegraphics[scale=0.318]{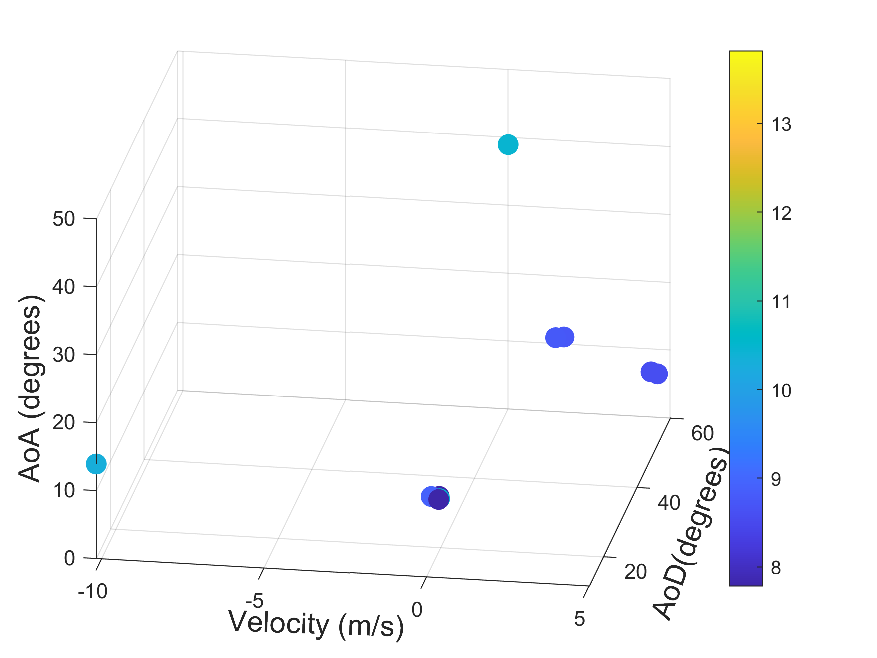}}
\caption{\textcolor{black}{(a) True target and (b) reconstructed target image of target RCS values in AD domain at $\text{SCNR}=-3\ \text{dB}$}}
\label{RadarImaging}
\end{figure}
the FIM $\mathbf {J}_{\mathcal {D}}$ can be expressed as
\begin{align}
    \mathbf {J}_{\mathcal {D}} = -\text {E}_{\left (\mathbf{y}, \boldsymbol{\beta} \right)}\left \lbrace{ \frac {\partial ^{2}\mathcal {L}\left ({\mathbf{y}\mid \boldsymbol{\beta};\mathbf {Z}}\right)}{\partial \boldsymbol{\beta}\partial \boldsymbol{\beta}^{H}}\!}\right \rbrace = \mathbf{\underline{D}}^H\mathbf{\underline{D}}.
\end{align}
The FIM component $\mathbf {J}_{\mathcal {P}}$ is given by
\begin{align}
    \mathbf {J}_{\mathcal {P}} = \mathbf{I}_{L} \otimes \boldsymbol{\Gamma}^{-1}.
\end{align}
Thus, the BCRB is given by
\begin{align}
    \text {E} \left\lbrace \vert \vert \mathbf {B} - \widehat{\mathbf {B}} \vert \vert _F^2 \right\rbrace &=  \text {E} \left\lbrace \vert \vert \boldsymbol{\beta}- \widehat{\boldsymbol{\beta}} \vert \vert _2^2 \right\rbrace \nonumber \\
    &=\text{Tr}\left\lbrace\mathbf {J}_{B}^{-1}\right\rbrace \nonumber \\
    &= \text{Tr}\left\lbrace\left(\mathbf{\underline{D}}^H\mathbf{\underline{D}} +\mathbf{I}_{L} \otimes \mathbf{Z}^{-1}\right)^{-1} \right\rbrace.
\end{align}
\begin{figure*}
	\centering
	\begin{minipage}{.87\columnwidth}
		\centering
		\includegraphics[width=\textwidth]{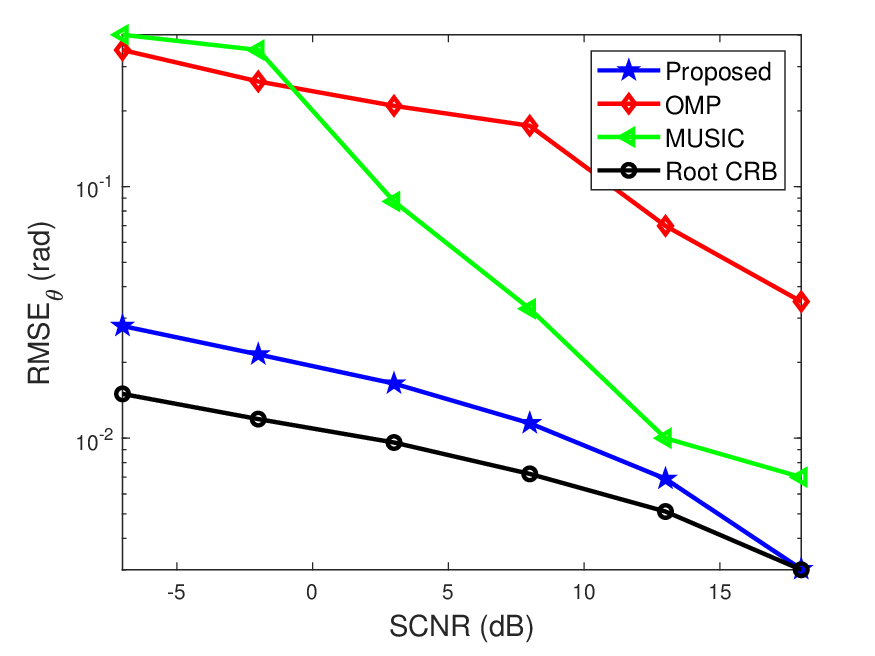}
		\label{AoDRMSE} \subcaption{}	
	\end{minipage}
	\begin{minipage}{.87\columnwidth}
		\centering		\includegraphics[width=\textwidth]{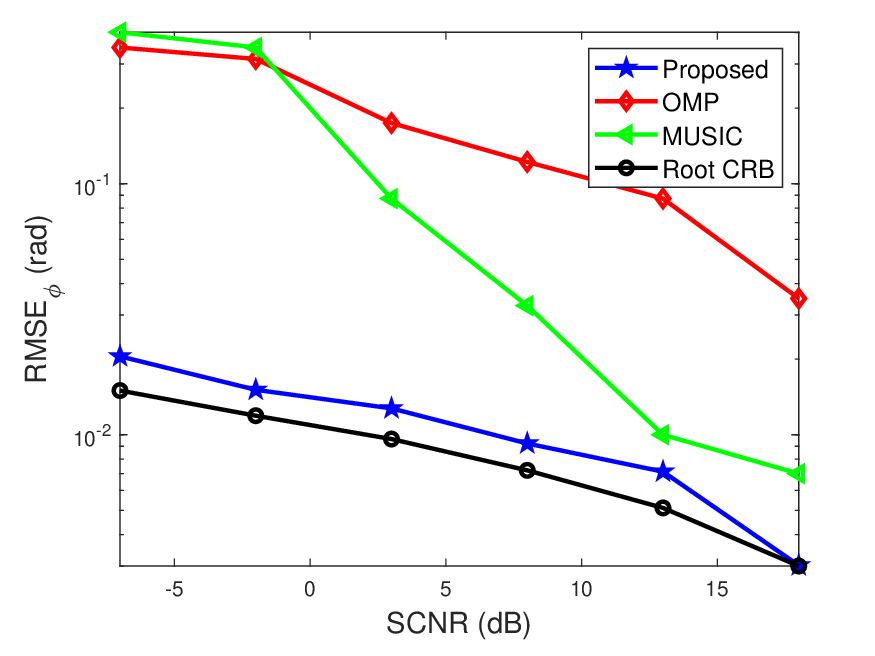}
		\label{AoARMSE} \subcaption{}	
	\end{minipage}
 \caption{\textcolor{black}{RMSE versus SCNR for estimation of (a) AoD and (b) AoA of targets for the proposed SBL, MUSIC and OMP algorithms}}
 \label{angleRMSE}
\end{figure*}
\begin{figure*}
	\centering
	\begin{minipage}{.67\columnwidth}
		\centering
		\includegraphics[width=\textwidth]{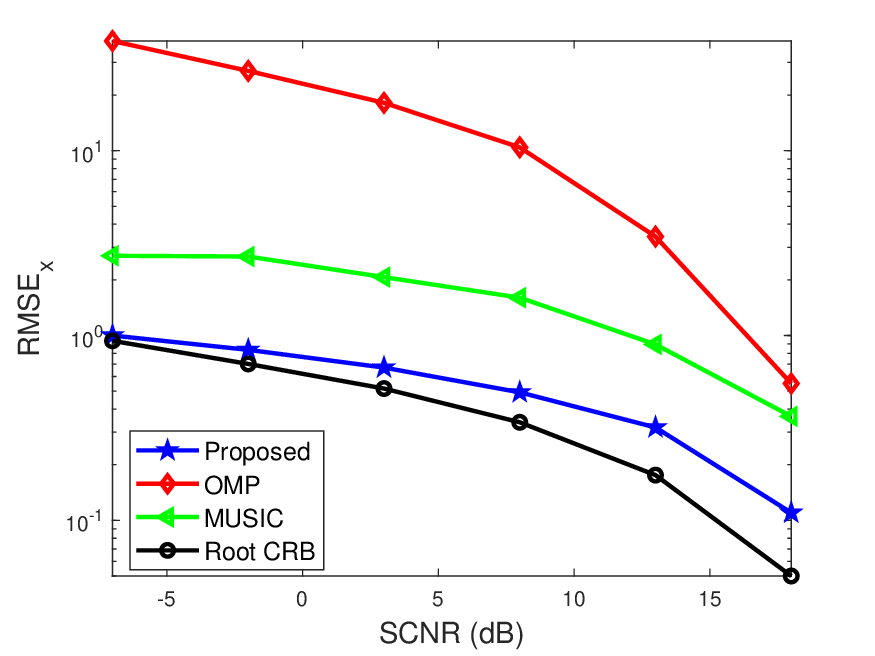}
		\label{XRMSE} \subcaption{}	
	\end{minipage}
	\begin{minipage}{.67\columnwidth}
		\centering		\includegraphics[width=\textwidth]{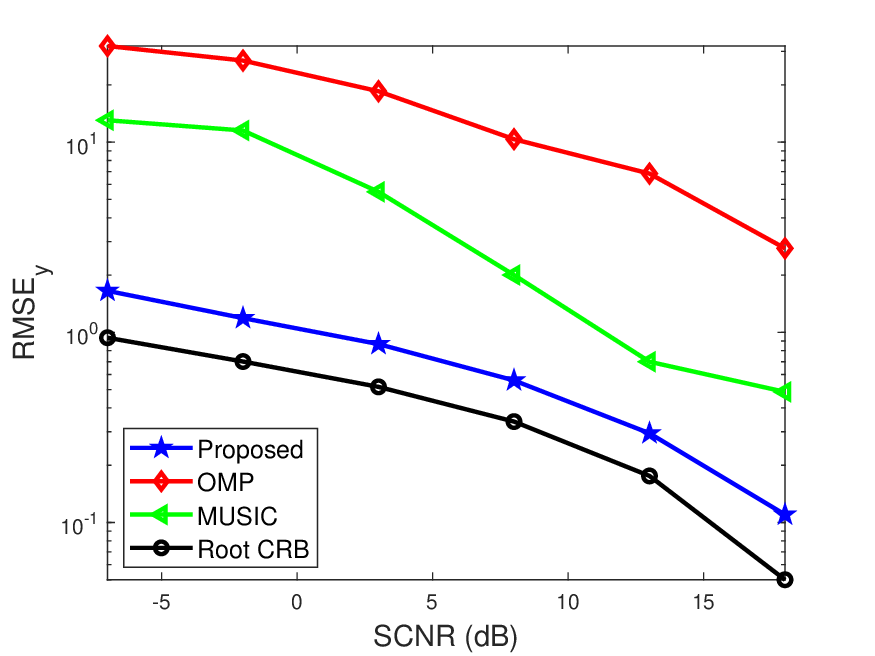}
		\label{YRMSE} \subcaption{}	
	\end{minipage}
 \begin{minipage}{.67\columnwidth}
		\centering		\includegraphics[width=\textwidth]{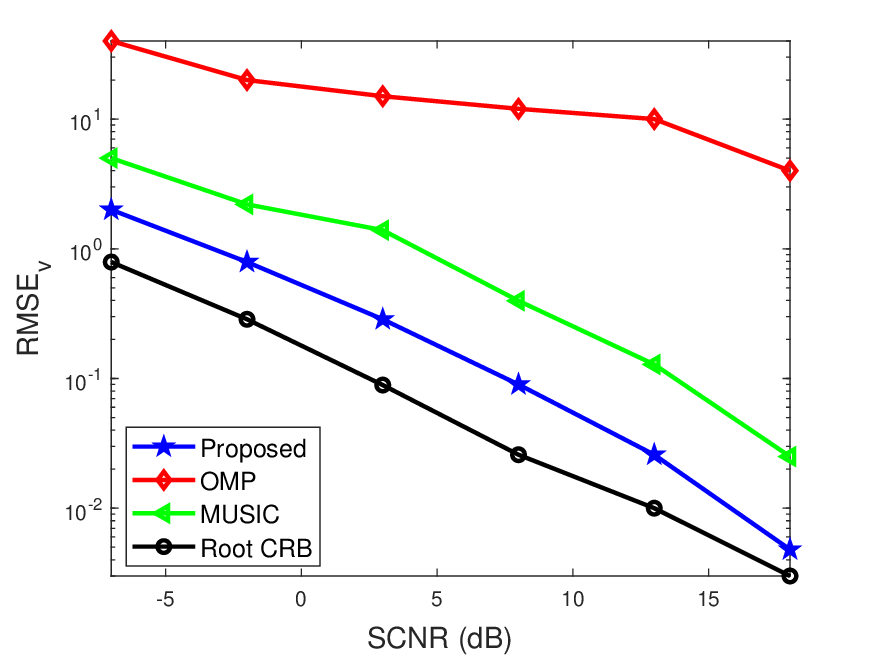}
		\label{VRMSE} \subcaption{}	
	\end{minipage}
 \caption{\textcolor{black}{RMSE versus SCNR for estimation of a) x-coordinate b) y-coordinate c) velocity of targets for the proposed SBL, MUSIC, and OMP algorithms}}
 \label{posRMSE}
\end{figure*}
\begin{table}[h]
\caption{{Simulation Setup}}
    \begin{tabular}{|l|r|}
 \hline
 \textbf{Parameter} & \textbf{Value} \\
 \hline
  Number of grids $G$  & 150  \\
\hline
 Number of targets $W$ & 3  \\
\hline
Number of clutter $N_{cl}$ & 2  \\
\hline
 Number of snapshots $L$ & 6 \\
\hline Number of transmit antennas $N_t$ & 6  \\
\hline
Number of transmit sub-pulses $U$ & 16  \\
\hline
Number of receive antennas $N_r$ & 6\\
\hline
Signal frequency $f_c$  & $30\ \text{GHz}$  \\
\hline
Sub-pulse duration $T_p$ & $40 \mu s$ \\
\hline
Polar coordinate of receiver $\left[R_r, \varphi_r \right]$ & $\left(95,0^o\right)$ \\
\hline
Polar coordinates of the targets $\left[R_w, \varphi_w \right]$  & $\left(20.66,52.64^o\right),$ \\$\left(\mathbf{u}_w,\,\!  w=1,2,\cdots, W\right)$ & $\left(25.11,54.73^o\right),$\\
$\ $ & $\left(51.13,12.08^o\right)$\\
\hline
Velocity of targets $(v_w)$ & $(5,2,-10) m/s$\\
\hline
    \end{tabular}
     \label{Tab:sim_par}
\end{table}
\section{Simulation Results} \label{results}
For characterizing the algorithm's target localization performance, we consider a mmWave MIMO radar system having the parameters specified in Table \ref{Tab:sim_par}.
Each element of the transmit signal matrix $\mathbf{X}\in \mathbb{C}^{N_t\times U} $ is a QPSK symbol of the form $\left\{\pm \sqrt{E_b} \pm j\sqrt{E_b}\right\}$, where $E_b$ is the energy per bit of the QPSK symbol.
The RCS coefficients of the targets and clutter are generated as i.i.d. samples of a zero-mean symmetric
complex Gaussian distribution with target and clutter variances of $\sigma_w^2 = 10 \text{dB}, \sigma_c^2 = 12 \text{dB}$, respectively. The noise samples are also i.i.d. zero-mean symmetric complex Gaussian of unit variance. The radar imaging scene spans the AoA and AoD region of $[0,\pi)$ and the Doppler shift spans in the region $[-1.5,1.5]$. 
Furthermore, let us define the target to clutter ratio (TCR) as the ratio of target variance to clutter variance. A high TCR represents a weaker clutter in comparison to the targets, and vice versa. The signal to clutter-plus-noise ratio (SCNR) is defined as $\text{SCNR}=E_b \frac{\sigma_w^2}{\sigma_c^2 + \sigma_n^2}$.

Fig. \ref{RadarImaging} juxtaposes the image of the radar scattering scene for the true target parameters to that of the respective estimated values. The color bars at the side denote the RCS coefficient. One can observe that the proposed SBL-based algorithm is capable of accurately identifying the locations of all the targets considered. Interestingly, it can be observed from the reconstructed image that the clutter components appear around the zero velocity, thereby, helping in the identification of the target parameter estimates from the clutter components.
\begin{figure}
	\centering
	\begin{minipage}{.87\columnwidth}
		\centering
		\includegraphics[width=\textwidth]{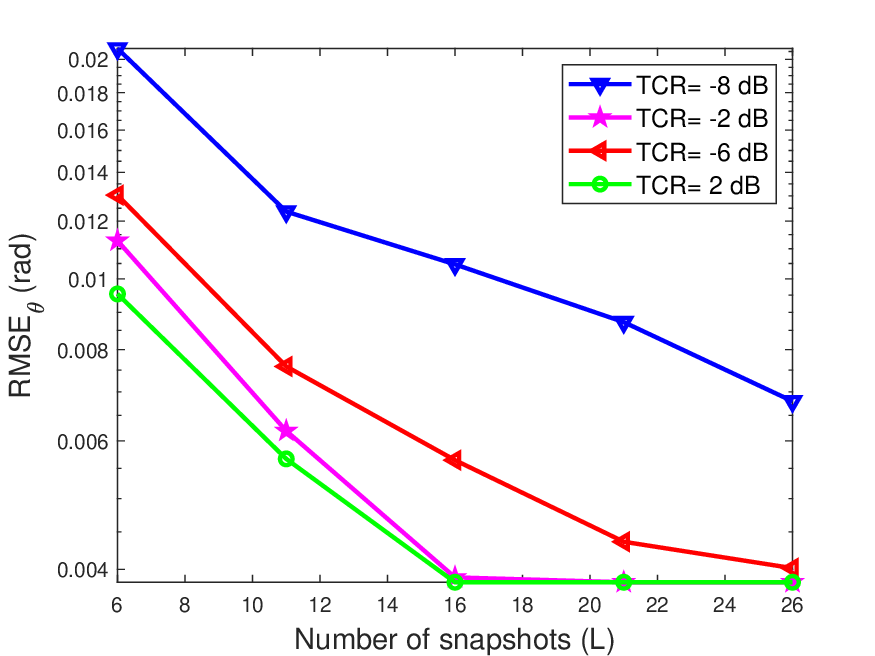}
		\label{AoDL} \subcaption{}	
	\end{minipage}
	\begin{minipage}{.87\columnwidth}
		\centering		\includegraphics[width=\textwidth]{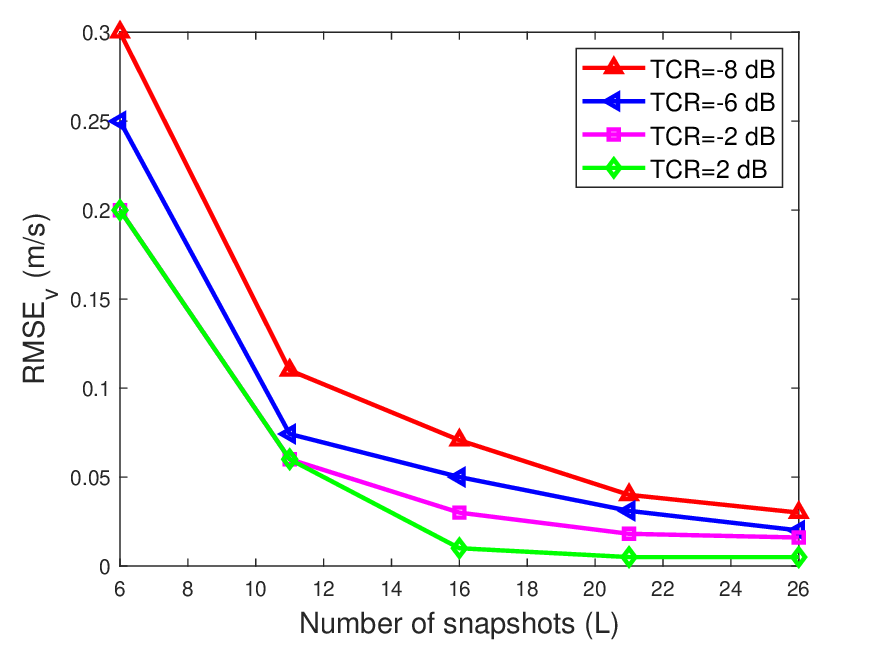}
		\label{velL} \subcaption{}	
	\end{minipage}
 \caption{RMSE of (a) AoD and (b) velocity versus number of snapshots $L$ for different values of TCR}
 \label{snapshot}
\end{figure}

We define the \textit{root mean squared error} (RMSE) metric to quantify the algorithm's parameter estimation accuracy as follows
\begin{align} \operatorname{RMSE}_{\eta}=&\sqrt{\frac{1}{M_c W} \sum _{i=1}^{M_c}\sum_{w=1}^{W}\left(\hat{\eta}_{w,i}-\eta _{w}\right)^{2}}, 
\end{align}
where we have $\eta_w \in \{ \theta _{w}, \phi _{w}, v _{w} \}$.

Fig. \ref{angleRMSE} (a) and (b) plot the RMSE of the estimated AoD and AoA values corresponding to different SCNR values both for the proposed SBL method and for the existing techniques. It can be observed that the RMSE values decrease upon increasing the SCNR. The proposed scheme outperforms both the existing CS-based OMP scheme \cite{5895106} and the subspace-based MUSIC scheme \cite{9512486} in the low SCNR regime, despite its lower dictionary overhead. The OMP scheme does not optimize the off-grid error between the true parameter and dictionary grid value and hence yields the poorest performance. The MUSIC algorithm deviates from the true values because it frequently suffers from rank deficiency of the covariance matrix for the case of correlated angles formed by closely spaced targets. \textcolor{black}{Moreover, as observed in the low SCNR regime of Fig. \ref{angleRMSE}, the MUSIC algorithm has a poor performance in comparison to OMP. This is due to the fact that at low SCNR values, the signal is heavily corrupted by the clutter-plus-noise components and leads to a lower signal subspace. By contrast, the OMP algorithm exploits the $3$D sparsity in the AD domain and yields a marginally improved performance over MUSIC.} Additionally, the MUSIC technique is prone to falsely identifying the peaks at clutter positions instead of the real targets, hence leading to poor performance. The proposed scheme also performs close to the root CRB approach, thus validating its supremacy.

\begin{figure}
\centering
\includegraphics[width=0.5\textwidth,scale=0.91]{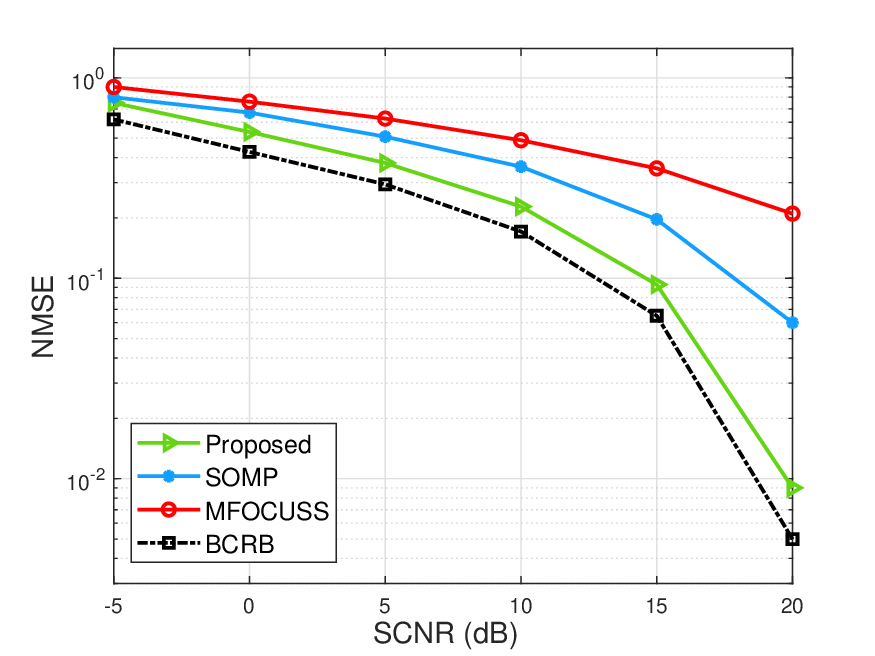}
\caption{NMSE versus SCNR for estimation of the RCS
coefficient matrix using the proposed SBL, SOMP and
MFOCUSS techniques} 
\label{rcs_nmse}
\end{figure}
\begin{figure}[h]
\includegraphics[width=0.5\textwidth,scale=1]{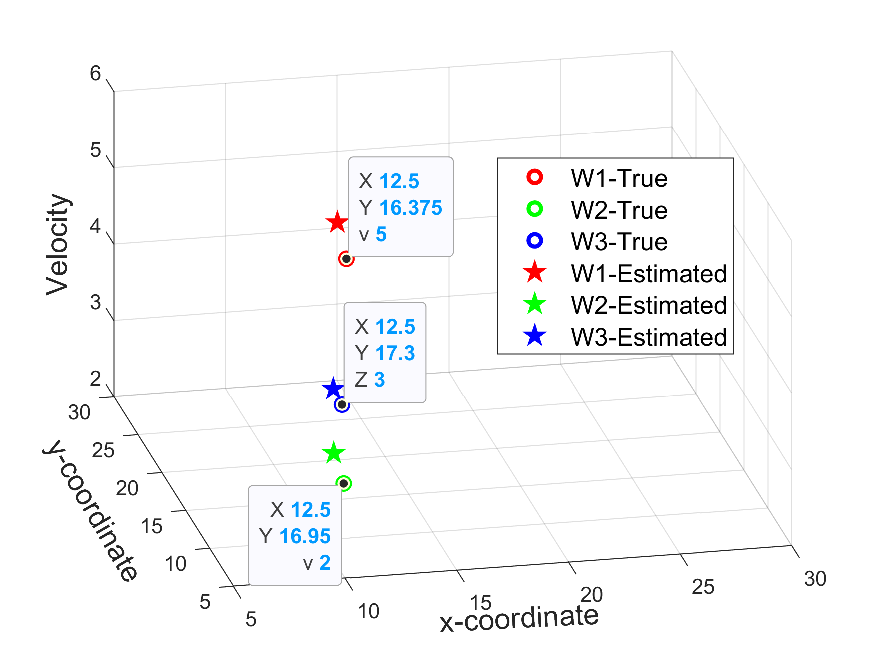}
\caption{\textcolor{black}{3D plot of the true and estimated target in terms of the x-coordinate, y-coordinate and velocity parameters}} 
\label{3dplot}
\end{figure}
\begin{figure}
\includegraphics[width=0.5\textwidth,scale=1]{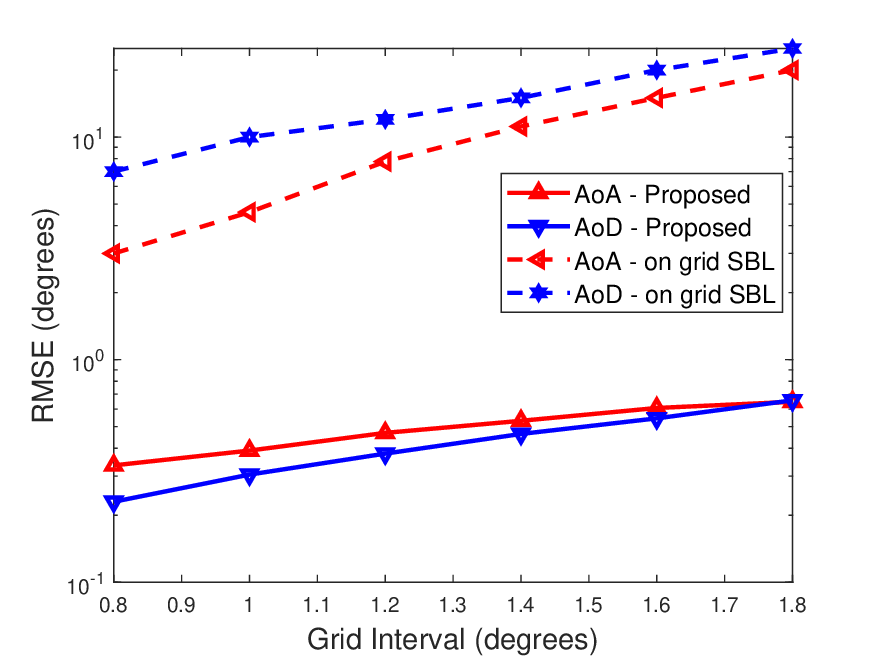}
\caption{RMSE versus dictionary grid interval for AoD and AoA estimation at $\text{SCNR} = -3\ \text{dB}$} 
\label{gridIntervalEffect}
\end{figure}
\begin{figure}
    \centering    \includegraphics[width=0.5\textwidth]{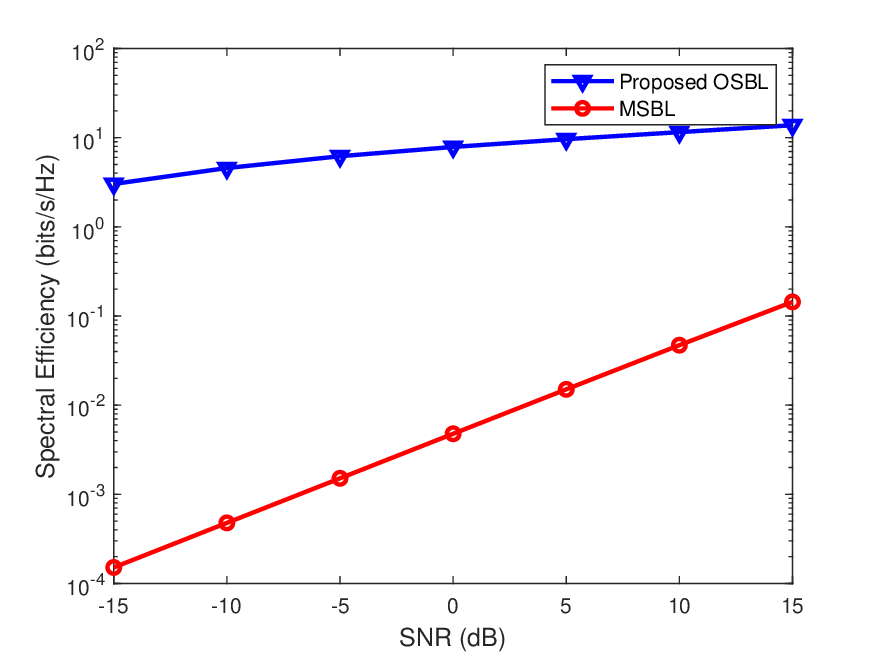}
    \caption{Spectral Efficiency comparison of the proposed positioning scheme and existing MSBL-based hybrid precoding and combining scheme in \cite{9179621}}
    \label{SEPlot}
\end{figure}
Fig. \ref{posRMSE} (a), (b) and (c) plot the RMSE of the estimated x-coordinate, y-coordinate and velocity, respectively, corresponding to different SCNR values both for the proposed SBL method and for the existing techniques. Since the position coordinates are directly derived from the angles as given in \eqref{locxy}, the location RMSE performance of the proposed scheme is superior to other schemes. Furthermore, the velocity RMSE performance seen in Fig. \ref{posRMSE} (c) is also as expected.

Fig. \ref{snapshot} depicts the AoD and velocity RMSE performance achieved versus the number of snapshots $L$ for different values of TCR. The RMSE performance improves upon increasing the number of snapshots. However, the performance of the proposed scheme saturates at approximately $L=22$ snapshots, which evidences its ability to provide accurate results with fewer snapshots. As a further benefit, the RMSE performance does not suffer at high TCR values, since it is capable of successfully isolating the clutter components from the targets. Both the above properties render it ideally suited for practical scenarios associated with a lower number of snapshots and high clutter.

Fig. \ref{rcs_nmse} portrays the \textit{normalized mean squared error} (NMSE) of the RCS coefficient matrix estimate $\widehat{\mathbf{B}}$ 
corresponding to different receive SCNR values. The NMSE can be defined as, NMSE $=\frac{1}{M_c}\sum _{i=1}^{M_c} \frac{ \Vert {{\widehat{\mathbf{B}}}} -{\mathbf{B}} \Vert _2^2 }{\Vert \widetilde{\mathbf{B}} \Vert _2^2}$, where $M_c$ is the number of Monte Carlo trials. The proposed SBL technique is observed to provide approximately $5$ dB and $10$ dB improvements over the OMP \cite{5895106} and the focal under-determined system solver (FOCUSS) \cite{558475} schemes, respectively. This is because the FOCUSS scheme is sensitive to the choice of the regularization parameter, whereas the OMP is affected both by the stopping criterion and the choice of the dictionary matrix, while the SBL performance is robust to these scenarios. Additionally, the NMSE of our algorithm is close to the BCRB, which validates its efficiency.

Fig. \ref{3dplot} illustrates the ability of the proposed super-resolution SBL technique
to resolve closely spaced targets. The target parameter values are as shown in Table \ref{tab2}.
\begin{table}
   \caption{True target parameter values for Fig. \ref{3dplot}}
    \label{tab2}
    \resizebox{\columnwidth}{!}{
 \begin{tabular}{ |c|c|c|c| }    
\hline
\textbf{Parameter} & \textbf{ Target 1 }& \textbf{ Target 2 }& \textbf{ Target 3 }\\
\hline
x-coordinate $(p_x)$ & $ 12.5$ m & $12.5$ m&  $12.5$ m\\
\hline
y-coordinate $(p_y)$ & $ 16.95$ m & $17.3$ m&  $16.375$ m\\
\hline
velocity $(v)$ & $2$ m/s & $3$ m/s& $5$ m/s\\
\hline
\end{tabular}}
 \end{table}
As observed from the figure, the proposed super-resolution SBL scheme is able to successfully distinguish multiple targets with a range resolution of $0.6$ m and velocity resolution of approximately $1$ m/s. 

Fig. \ref{gridIntervalEffect} depicts the estimation performance of both the proposed scheme and of the on-grid SBL, where the latter is configured by changing the granularity of the grid points chosen. It can be readily deduced that the RMSE performance of our scheme conceived is robust, leading to successful localization of the target. However, the on-grid SBL performance degrades significantly in comparison to the proposed off-grid SBL for coarser grids. The on-grid SBL requires approximately twice the dictionary size compared to the off-grid SBL in order to achieve the same RMSE performance. These results collectively imply that the proposed algorithm offers a higher parameter estimation accuracy at a substantially reduced complexity, rendering it ideally suited for practical implementation.

To substantiate the proposed scheme's contributions in communication systems, we have presented the spectral efficiency (SE) comparison of the hybrid precoding and combining using the proposed positioning scheme with existing multiple measurement vector (MMV)-based sparse Bayesian learning (MSBL)-based hybrid precoding and combining scheme in \cite{9179621}. We assume that the targets act as scatterers in mmWave MIMO channel $\mathbf{H}=\sum_{w=1}^W \beta_w \mathbf{d}_w \mathbf{c}_w^T e^{j\omega_w} \in \mathbb{C}^{M_r \times M_t}$. Consider a setup with $N_s$ transmit symbols and $N_{\text{RF}}$ RF chains. Once one obtains the path state information of the targets (scatterers) such as AoA $\phi$, AoD $\theta$, Doppler shift $\omega$ and RCS coefficients(channel coefficients) $\beta$, the RF precoder $\mathbf{F}_{\text{RF}} \in \mathbb{C}^{M_t \times N_{\text{RF}}}$ and RF combiner $\mathbf{W}_{\text{RF}} \in \mathbb{C}^{M_r \times N_{\text{RF}}}$ can be designed by choosing the dominant transmit and receive steering matrices, respectively. Specifically, we arrange the estimated RCS coefficients (channel coefficients) in a decreasing order, $\vert \widehat{\beta}_{w_1} \vert \geq \vert \widehat{\beta}_{w_2} \vert \geq \cdots \geq \vert \widehat{\beta}_{w_W} \vert $. Thereafter,
the RF precoder $\mathbf{F}_{\text{RF}}$ and RF combiner
 $\mathbf{W}_{\text{RF}}$ are designed as
\begin{align}
    \mathbf{F}_{\text{RF}} = [\mathbf{c}({w_1}), \mathbf{c}({w_2}), \cdots, \mathbf{c}(w_{N_{\text{RF}}})] , \\
     \mathbf{W}_{\text{RF}} = [\mathbf{d}({w_1}), \mathbf{d}({w_2}), \cdots, \mathbf{d}(w_{N_{\text{RF}}})] .
\end{align} 
Upon performing singular value decomposition (SVD) of the effective channel $\widetilde{\mathbf{H}} =  \mathbf{W}_{\text{RF}}^H \mathbf{H} \mathbf{F}_{\text{RF}} = \mathbf{U}\boldsymbol{\Sigma}\mathbf{V}^H$, the baseband precoder $\mathbf{F}_{\text{BB}}$ and combiner $\mathbf{W}_{\text{BB}}$ can be designed as $\mathbf{F}_{\text{BB}} = \mathbf{V}, \mathbf{W}_{\text{BB}} = \mathbf{U}$. One can observe from Fig. \ref{SEPlot} a significant SE performance gain using the proposed positioning scheme. The improved SE performance is attributed to the fact that the RF precoder $\mathbf{F}_{\text{RF}} $ and RF combiner $\mathbf{W}_{\text{RF}} $ designed from the  proposed positioning scheme provides significant beamforming gain compared to the scheme proposed in \cite{9179621} which does not account for the off-grid errors.
\section{Conclusions} \label{conclusion}
A joint angle-velocity-RCS coefficient estimation algorithm was devised for the localization of multiple targets in a bistatic mmWave MIMO radar system operating in the face of clutter. The proposed scheme exploits the 3D-sparsity of the target in the AD domain for jointly estimating the target parameters, such as its velocity, angles and RCS coefficients. In order to mitigate the estimation errors arising due to grid mismatch, an innovative SBL algorithm was formulated for recursively updating the parameter grid points in each iteration, thereby ensuring that they approach the true target parameter values upon convergence. The AD-domain representation also helps us to distinguish the estimated target parameters from the clutter. We also derived the CRBs for target parameter estimation to benchmark the estimation performance. Our simulation results validated the superiority of the proposed framework compared to the existing solutions. Possible future extension of this work might consider the introduction of reconfigurable intelligent surface (RIS) to tackle the challenging problem of target localization in the blind zone, wherein the transmitted signal is unable to reach the target due to obstacles in the LoS path.  
\begin{appendices}
\section{Proofs of \eqref{z_update_final}, \eqref{et_update_final}, \eqref{er_update_final}, \eqref{ew_update_final}} \label{proof}
We present the detailed derivations for \eqref{z_update_final}, \eqref{et_update_final}, \eqref{er_update_final}, \eqref{ew_update_final}.
    \subsection{Update for $\mathbf{z}$ in \eqref{z_update_final}}
The objective function in \eqref{z_update} can be rewritten as follows after ignoring the independent terms 
\begin{align}
  &  \mathcal{L}\big(\mathbf{z}, \boldsymbol{\epsilon}_t^{(k)}, \boldsymbol{\epsilon}_r^{(k)}, \boldsymbol{\epsilon}_{\omega}^{(k)} \vert \cdot,\cdot,\cdot,\cdot,\cdot,\cdot\big) = \nonumber \\
  &  \int{p \left(\mathbf{B}\vert \mathbf{Y},\mathbf{z}^{(k)}, \boldsymbol{\epsilon}_t^{(k)}, \boldsymbol{\epsilon}_r^{(k)}, \boldsymbol{\epsilon}_{\omega}^{(k)} \right) \text{ln}\  p(\mathbf{B\vert z}) d\mathbf{B}} \nonumber \\
    & + \int{p \left(\mathbf{B}\vert \mathbf{Y},\mathbf{z}^{(k)}, \boldsymbol{\epsilon}_t^{(k)}, \boldsymbol{\epsilon}_r^{(k)}, \boldsymbol{\epsilon}_{\omega}^{(k)} \right) \text{ln}\  p(\mathbf{z}) d\mathbf{B}} \nonumber \\
    & = (a+L) \sum_{g=1}^G \text{ln} z_g - b \sum_{g=1}^G z_g - \text{Tr} \left( \sum_{l=1}^L \Upsilon_l \left( \boldsymbol{\epsilon}_t^{(k)}, \boldsymbol{\epsilon}_r^{(k)}, \boldsymbol{\epsilon}_{\omega}^{(k)}\right) \right),
 \end{align}
where we have $\Upsilon_l \left( \boldsymbol{\epsilon}_t^{(k)}, \boldsymbol{\epsilon}_r^{(k)}, \boldsymbol{\epsilon}_{\omega}^{(k)}\right) = \boldsymbol{\mu}_l \boldsymbol{\mu}^H_l + \boldsymbol{\Sigma}$. Upon differentiating the above expression with respect to $z_g$ and equating it to zero, we obtain \eqref{z_update_final}. 
\subsection{Update for $\boldsymbol{\epsilon}_t$ in \eqref{et_update_final}}
After ignoring independent terms, the objective function in \eqref{et_update} can be rewritten as follows
\begin{align}
  &  \mathcal{L}\big(\mathbf{z}^{(k+1)}, \boldsymbol{\epsilon}_t, \boldsymbol{\epsilon}_r^{(k)}, \boldsymbol{\epsilon}_{\omega}^{(k)} \vert \cdot,\cdot,\cdot,\cdot\big) = \nonumber \\
  & \int{p \left(\mathbf{B}\vert \mathbf{Y}, \cdot,\cdot,\cdot,\cdot  \right) \text{ln}\  p\big(\mathbf{Y\vert B},\mathbf{z}^{(k+1)},  \boldsymbol{\epsilon}_t, \boldsymbol{\epsilon}_r^{(k)},\boldsymbol{\epsilon}_{\omega}^{(k)}\big) d\mathbf{B} } \nonumber \\
      &= - \int{p \left(\mathbf{B}\vert \mathbf{Y}, \cdot,\cdot,\cdot,\cdot \right)} \sum_{l=1}^L \big({\mathbf{y}}_l -(\boldsymbol{\Psi}_{t}+\boldsymbol{\Xi}_t \text{diag}(\boldsymbol{\epsilon}_t))\boldsymbol{\beta}_l  \big)^H \nonumber \\
      & \times \big({\mathbf{y}}_l - (\boldsymbol{\Psi}_{t}+\boldsymbol{\Xi}_t \text{diag}(\boldsymbol{\epsilon}_t))\boldsymbol{\beta}_l \big) d\mathbf{B}\nonumber \\
    & = -  \sum_{l=1}^L \left({\mathbf{y}}_l - \widetilde{\boldsymbol{\Psi}}_t\boldsymbol{\mu}_l  \right)^H  \left({\mathbf{y}}_l - \widetilde{\boldsymbol{\Psi}}_t \boldsymbol{\mu}_l  \right) - L\text{Tr}\bigg(\widetilde{\boldsymbol{\Psi}}_t  \boldsymbol{\Sigma} \boldsymbol{\mu}_l \widetilde{\boldsymbol{\Psi}}_t^H \bigg) \nonumber \\
    & = -\boldsymbol{\epsilon}_t^H\left( \Re\left\{ \boldsymbol{\Xi}_t^H\boldsymbol{\Xi}_t \odot (\mathbf{U}\mathbf{U}^H + L\boldsymbol{\Sigma})  \right\} \right)\boldsymbol{\epsilon}_t + 2\mathbf{p}_t^H\boldsymbol{\epsilon}_t +c,
 \end{align}
where 
$\widetilde{\boldsymbol{\Psi}}_t =  \boldsymbol{\Psi}_{t}+\boldsymbol{\Xi}_t \text{diag}\left(\boldsymbol{\epsilon}_t\right)$ and $c$ is a constant term.
Upon differentiating the above expression with respect to the offset vector $\boldsymbol{\epsilon}_t $ and equating it to zero, we obtain \eqref{et_update_final}.
\subsection{Update for $\boldsymbol{\epsilon}_r$ in \eqref{er_update_final}}
Upon ignoring the independent terms, the objective function in \eqref{er_update} can be rewritten as follows
\begin{align}
  &  \mathcal{L}\big(\mathbf{z}^{(k+1)}, \boldsymbol{\epsilon}_t^{(k+1)}, \boldsymbol{\epsilon}_r, \boldsymbol{\epsilon}_{\omega}^{(k)} \vert \cdot,\cdot,\cdot\big) = \nonumber \\
  & \int{p \left(\mathbf{B}\vert \mathbf{Y}, \cdot,\cdot,\cdot,\cdot  \right) \text{ln}\  p\big(\mathbf{Y\vert B},\mathbf{z}^{(k+1)}, \boldsymbol{\epsilon}_t^{(k+1)}, \boldsymbol{\epsilon}_r,\boldsymbol{\epsilon}_{\omega}^{(k)}\big) d\mathbf{B}}  \nonumber \\
    & = -  \sum_{l=1}^L \big({\mathbf{y}}_l - \widetilde{\boldsymbol{\Psi}}_{r}\boldsymbol{\mu}_l  \big)^H   \big({\mathbf{y}}_l - \widetilde{\boldsymbol{\Psi}}_{r}  \boldsymbol{\mu}_l  \big)-  L\text{Tr}\bigg( \widetilde{\boldsymbol{\Psi}}_{r}\boldsymbol{\Sigma} \widetilde{\boldsymbol{\Psi}}_{r}^H \bigg) \nonumber \\
    & = -\boldsymbol{\epsilon}_r^H\left( \Re\left\{ \boldsymbol{\Xi}_r^H \boldsymbol{\Xi}_r \odot (\mathbf{U}\mathbf{U}^H + L\boldsymbol{\Sigma})  \right\} \right)\boldsymbol{\epsilon}_r + 2\mathbf{p}_r^H\boldsymbol{\epsilon}_r +c,
 \end{align}
where $\widetilde{\boldsymbol{\Psi}}_{r} = \boldsymbol{\Psi}_{r}+\boldsymbol{\Xi}_r \text{diag}(\boldsymbol{\epsilon}_r)$, and $c$ is a constant term.
Eq. \eqref{er_update_final} is obtained after differentiating the above expression with respect to the offset vector $\boldsymbol{\epsilon}_r $ and equating it to zero.
\subsection{Update for $\boldsymbol{\epsilon}_{\omega}$ in \eqref{ew_update_final}}
After ignoring the independent terms, the objective function in \eqref{ew_update} can be rewritten as follows
\begin{align}
  &  \mathcal{L}\big(\mathbf{z}^{(k+1)}, \boldsymbol{\epsilon}_t^{(k+1)}, \boldsymbol{\epsilon}_r^{(k+1)}, \boldsymbol{\epsilon}_{\omega} \vert \cdot,\cdot,\cdot,\cdot\big) = \nonumber \\
  & \int{p \left(\mathbf{B}\vert \mathbf{Y}, \cdot,\cdot,\cdot,\cdot\right) \text{ln}\  p\big(\mathbf{Y\vert B},\mathbf{z}^{(k+1)}, \boldsymbol{\epsilon}_t^{(k+1)}, \boldsymbol{\epsilon}_r^{(k+1)},\boldsymbol{\epsilon}_{\omega}\big) d\mathbf{B}}  \nonumber \\
    & = -  \sum_{l=1}^L \big({\mathbf{y}}_l - \widetilde{\boldsymbol{\Psi}}_{\omega}\boldsymbol{\mu}_l  \big)^H    \big({\mathbf{y}}_l - \widetilde{\boldsymbol{\Psi}}_{\omega}\boldsymbol{\mu}_l  \big)-  L\text{Tr}\bigg(\widetilde{\boldsymbol{\Psi}}_{\omega} \boldsymbol{\Sigma} \widetilde{\boldsymbol{\Psi}}_{\omega}^H \bigg) \nonumber \\
    & = -\boldsymbol{\epsilon}_{\omega}^H\left( \Re\left\{ \boldsymbol{\Xi}_{\omega}^H \boldsymbol{\Xi}_{\omega} \odot (\mathbf{U}\mathbf{U}^H + L\boldsymbol{\Sigma})  \right\} \right)\boldsymbol{\epsilon}_{\omega} + 2\mathbf{p}_{\omega}^H\boldsymbol{\epsilon}_{\omega} +c,
 \end{align}
where $\widetilde{\boldsymbol{\Psi}}_{\omega} = \boldsymbol{\Psi}_{\omega}+\boldsymbol{\Xi}_{\omega} \text{diag}(\boldsymbol{\epsilon}_{\omega})$, and $c$ is a constant term.
Upon differentiating the above expression with respect to the offset vector $\boldsymbol{\epsilon}_{\omega} $ and equating it to zero, we obtain \eqref{ew_update_final}.
\end{appendices}
\bibliographystyle{IEEEtran}
\bibliography{References}
\end{document}